\documentclass[a4paper,fleqn,usenatbib]{mnras}

\usepackage[T1]{fontenc}
\usepackage{ae,aecompl}

\usepackage{graphicx}
\usepackage{times}
\usepackage{array}
\usepackage{stfloats}
\usepackage{amsmath}
\usepackage{xspace}
\usepackage{txfonts}
\usepackage{ulem}                   
\usepackage{hyperref}
\usepackage{amssymb}
\usepackage[dvipsnames]{xcolor}

\graphicspath{{./images/}}

\newcommand{\hMpc}{h^{-1}\,{\rm Mpc}}

\newcommand{\Rlc}{R_\mathrm{lc}}
\newcommand{\lmax}{l_\mathrm{max}}
\newcommand{\tgeom}{\theta_\mathrm{geom}}
\newcommand{\adyn}{\alpha_\mathrm{dyn}}
\newcommand{\Lbox}{L_\mathrm{box}}
\newcommand{\std}{\textit{std}\xspace}
\newcommand{\dzs}{\textit{dzs}\xspace}

\newcommand{\gadget}{\texttt{PGADGET-3}\xspace}

\newcommand{\eg}{e.g.\xspace}
\newcommand{\ie}{i.e.\xspace}
\newcommand{\citenp}[1]{\citeauthor{#1} \citeyear{#1}}

\title[Dynamic Zoom Simulations]{Dynamic Zoom Simulations: a fast, adaptive algorithm for simulating lightcones}

\author[E. Garaldi et al.]{
Enrico Garaldi,$^{1}$\thanks{egaraldi@mpa-garching.mpg.de} 
Matteo Nori,$^{2,3,4}$
and Marco Baldi$^{2,3,4}$
\\
$^{1}$Max Planck Institute for Astrophysics, Karl-Schwarzschild-Stra{\ss}e 1, 85741 Garching, Germany\\
$^{2}$Dipartimento di Fisica e Astronomia, Universit\'a di Bologna, via Piero Gobetti 93/2, 40129 Bologna, Italy\\
$^{3}$INAF - Osservatorio di Astrofisica e Scienza dello Spazio, via Piero Gobetti 93/3 1, 40129 Bologna, Italy\\
$^{4}$INFN - Sezione di Bologna, viale Berti Pichat 6/2, 40127 Bologna, Italy\\
}

\date{Accepted XXX. Received YYY; in original form ZZZ}

\pubyear{}

\begin{document}
\label{firstpage}
\pagerange{\pageref{firstpage}--\pageref{lastpage}}
\maketitle

\begin{abstract}
The advent of a new generation of large-scale galaxy surveys is pushing cosmological numerical simulations in an uncharted territory. The simultaneous requirements of high resolution and very large volume pose serious technical challenges, due to their computational and data storage demand. In this paper, we present a novel approach dubbed Dynamic Zoom Simulations --~or DZS~-- developed to tackle these issues. Our method is tailored to the production of lightcone outputs from N-body numerical simulations, which allow for a more efficient storage and post-processing compared to standard comoving snapshots, and more directly mimic the format of survey data. In DZS, the resolution of the simulation is dynamically decreased outside the lightcone surface, reducing the computational work load, while simultaneously preserving the accuracy inside the lightcone and the large-scale gravitational field. We show that our approach can achieve virtually identical results to traditional simulations at half of the computational cost for our largest box. We also forecast this speedup to increase up to a factor of $5$ for larger and/or higher-resolution simulations. We assess the accuracy of the numerical integration by comparing pairs of identical simulations run with and without DZS. Deviations in the lightcone halo mass function, in the sky-projected  lightcone, and in the 3D matter lightcone always remain below $0.1$\%. In summary, our results indicate that the DZS technique may provide a highly-valuable tool to address the technical challenges that will characterise the next generation of large-scale cosmological simulations.
\end{abstract} 


\section{Introduction}
\label{sec:intro}

Numerical simulations play a pivotal role in cosmology and astrophysics. Their importance surged in the last few decades, establishing them as one of the primary investigation tools in these disciplines. However, cosmological observations of  ever-growing precision require increasingly accurate --~and thus expensive~-- simulations to be interpreted. 
Forthcoming instruments like \textit{Euclid} \citep{EUCLID}, WFIRST
\citep{WFIRST}, the Vera C. Rubin Observatory \citep[formerly known as LSST, ][]{LSST} and DESI \citep{DESI1,DESI2} will provide an unprecedented amount of data, outperforming all previous observational campaigns in terms of both quality and covered volume. Their interpretation therefore demands a new generation of numerical simulations comprising volumes of the universe comparable in size with those covered by such surveys and --~simultaneously~-- of high physical fidelity.

This upcoming new era of numerical simulations faces a series of technical challenges. First and foremost, 
the total time
 required for their completion increases more than linearly with 
both resolution and volume. In addition, the information stored 
per resolution element is hardly reducible, entailing that the 
memory requirements of such simulations keep increasing, in contrast with the trend in high-performance computing facilities to decrease the amount of available memory --~both on the machine and the CPU level.

Hence, even on the top-ranked supercomputing infrastructures currently available, the most advanced simulations can easily necessitate of large fractions of the entire supercomputer for very long time. For instance, the TNG50 run of the Illustris-TNG simulation project \citep[which follows gravitational, magnetic, and hydrodynamical interactions,][]{TNG50-Dylan, TNG50-Annalisa} required approximately $128$ TB of memory for over $90$ million CPU-hours. Similarly, the Euclid flagship simulation \citep{pkdgrav-euclid-flagship} employed a similar amount of memory for more than $4$ million CPU-hours to follow the evolution of a Dark Matter-only universe. 

Even 
if the aforementioned issues 
were taken care of, the community 
would be faced with the technical challenge of storing the data produced by such large simulations. The typical approach 
is to output time slices of the whole simulation volume (\ie saving the properties of all simulated particles  
at a given cosmic time in a so-called snapshot) and then, in post-processing, inspect them as a time series. When trying to mimic the survey view of the Universe (where images of structures at different cosmic times coexist), snapshots are combined in order to create lightcones-like data \citep[\eg][]{Hollowed2019}. This can be done either with a piecewise-constant approximation or interpolating the particle position between adjacent outputs. In both cases, a large number of time slices are necessary in order to reach the degree of fidelity required by modern galaxy survey data, with only a small fraction of the particles in each snapshot concurring to the lightcone reconstruction. 
However, such procedure would require a prohibitively-large amount of storing space  
in the case of the most advanced simulations.

This technical challenge prompted some simulators to abandon the traditional snapshot format in favor of lightcone outputs (pioneered by \citenp{Evrard+2002}, and more recently employed \eg in \citenp{OnionUniverse}, in the Euclid flagship simulation presented in \citenp{pkdgrav-euclid-flagship}, and in the $768 \, \hMpc$ run of the simulations in \citenp{MICE-lc}). In this approach, lightcones are produced on the fly by saving to disk only the particles that are, at any given redshift $z$, in a thin spherical shell with outer radius equal to the lightcone radius\footnote{The ligthcone radius is defined as the comoving distance $\chi (z)$ along a null geodesic stemming from the observer at $z=0$ and connecting it to an event situated at redshift $z$.} $\Rlc (z)$, centered on the observer. They represent the only output of the simulation (possibly complemented by few traditional snapshots). This enables simulations to reach a much more fine-grained discretisation of the lightcone while saving a large amount of storage space, as only particles 
belonging to the lightcone 
are stored.

While tackling the issue of storing and post-processing simulation data, the use of lightcone outputs does not ease the challenging run-time memory and computational power requirements of such massive numerical enterprises. Recently \citet[][L17 hereafter]{Llinares17} proposed a new simulation method dubbed the \textit{Shrinking Domain Framework} (SDF) as a possible solution to these problems. Such approach exploits the fact that the lightcone radius $\Rlc$ decreases with time and, simultaneously, most of the computational effort is invested at late times, as a consequence of increased clustering of matter. In the SDF, particles outside of the lightcone (\ie more distant than $\Rlc$ from the observer) are discarded from the numerical integration, thereby speeding up the simulation since the number of resolution elements to be evolved decreases with time as the lightcone volume shrinks onto the observer. However, the performance improvement obtained with such a radical modification of the simulation structure does not come for free: the price to be paid for this faster algorithm is the loss of periodicity of the system, which is a fundamental assumption for many numerical gravity solvers. As a consequence, the nature of the gravitational solver changes, transforming the Poisson equation into an equation that accounts for the finite information propagation speed\footnote{The equation used in L17 is not the correct one and should be replaced by the trace of the Einstein's equation. Preliminary tests show no change in results or performance (Llinares, private communication).}. Additionally, density perturbations (and therefore gravitational interactions) on scales larger than $\Rlc$ are effectively ignored. While the latter would not have any impact inside the lightcone in a fully-relativistic treatment of gravity, this is not the case in the Newtonian approximation typically enforced in many cosmological simulation codes.

A totally different approach to the problem of reducing the computational cost of large cosmological simulations has been proposed in \citet{COLA}. Their method, dubbed COmoving Lagrangian Acceleration (COLA), solves the large-scale motion using the second order Lagrangian perturbation theory (LPT), while the small-scale motion is integrated using a N-body code. This approach allows to sacrifice accuracy on small scales by forcing large timesteps to gain orders of magnitude in simulation speed. As a result, the COLA method approximates the large-scale matter distribution very well, but fails to match the internal structure of Dark Matter haloes, which is crucial for producing realistic galaxies. Additionally, this method does not allow the treatment of baryons \citep[although it has been proposed to use a gradient-based method to mimic the output of hydrodynamical simulations, at the cost of introducing nuisance parameters, see \eg][]{COLAbaryons}. In a further development of the COLA algorithm dubbed spatial COLA \citep[or sCOLA,][]{sCOLA}, small- and large-scales modes are de-coupled by computing the latter via LPT and superimposing them to the former, whose evolution is solved using an $N$-body integrator. This approach can also be used to tile a number of independent sCOLA simulations to produce a larger synthetic universe \citep{Leclercq+2020}, effectively removing any overhead due to inter-task communication. However, it does not reduce the CPU time required to integrate small-scale motion, but rather provides an efficient way to distribute it across many computing tasks. Lastly, it can be applied only when an entire non-linear region can be contained in a single computing task, \ie when particle crossing the tile boundary can be ignored.

In this paper we present a new method dubbed Dynamic Zoom Simulation (DZS) devised to overcome 
the technical challenges faced by modern simulations. Similarly to the SDF, the algorithm is designed to focus computational effort only in the regions of the simulation domain that are necessary for building a lightcone with given depth. However, in order to retain the large-scale gravity modes and render the implementation simpler and suitable to virtually all simulation codes available, we do not discard particles at distance $d > \Rlc$ from the lightcone center. Instead, we progressively decrease their resolution by merging together multiple particles, according to an adjustable set of rules. This approach beats down the computational cost and, simultaneously, does not require any change in the gravity solver.

The paper is organized as follows. In Sec.~\ref{sec:dzs} we describe the details of the DZS algorithm and its implementation in the 
popular simulation code \gadget \citep[a previous version of which has been described in ][]{gadget1,gadget2}. Then, we describe a suite of numerical tests (Sec.~\ref{sec:tests}) exploring the accuracy of the DZS algorithm. Its performance is studied in Sec.~\ref{sec:performance}. We provide concluding remarks in Sec. \ref{sec:conclusions}.

\section{Dynamic Zoom Simulations}
\label{sec:dzs}

\begin{figure*}
    \centering
    \includegraphics[width=\textwidth]{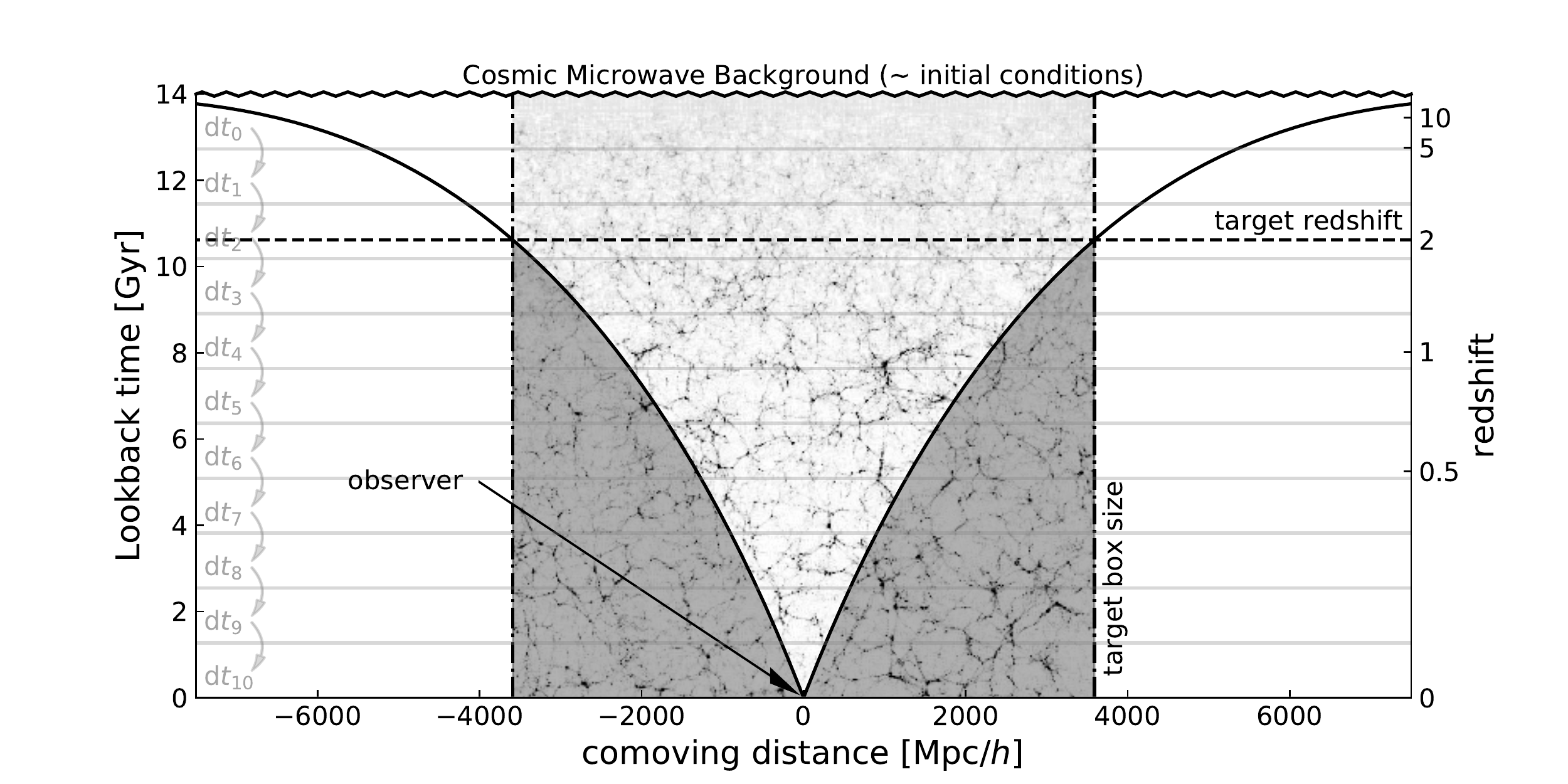}
    \caption{Space-time diagram (in 1+1 dimensions). The black solid lines show the lightcone for an observer placed at the origin assuming a \citet{Planck2015_cosmology} cosmology. The maximum (`target') redshift of the simulated lightcone (horizontal dashed line) corresponds to a minimum (`target') box size (vertical dot-dashed line). In normal simulations aiming at producing a lightcone output, the entire box is simulated from the initial conditions (undulating line on top) down to redshift zero. This is illustrated by the image in the background (representing a simulated cosmological density field in a $512 \, \hMpc$ box, and used only for displaying purposes). Usually, the initial conditions are evolved until redshfit $z \gtrsim 50$ using linear theory, in order to save computational time. For the sake of simplicity, we have disregarded such step in the scheme. The grey-shaded areas highlight regions of space-time that are simulated but discarded, as not part of the lightcone. Finally, the thin light-gray horizontal lines exemplify the integration procedure adopted in simulations, where the entire box is advanced in time by a timestep $\mathrm{d}t$.}
    \label{fig:scheme_lc}
\end{figure*}

In recent years large cosmological simulations shifted from the (historical) snapshot outputs to lightcone ones. This shift has been driven both by the aforementioned technical challenges of storing a large number of snapshots for increasingly large simulations, and by the fact that galaxy surveys data --~\ie the main target of such large simulations~-- naturally come in the form of a lightcone. However, the internal structure of N-body codes has not evolved to reflect this change. In fact, simulation algorithms still commonly integrate the motion of all particles in the whole simulation volume, including those situated at distances $d > \Rlc$ from the observer. 
Hence, a substantial amount of computation time is spent on regions of the simulation box that are then discarded in the output process. At low redshift\footnote{The relevant redshift depends chiefly on the box size $\Lbox$, which determines the redshift $\Tilde{z}$ at which the lightcone encloses a volume significantly smaller than the simulation box, \ie $(4/3) \pi \Rlc(\Tilde{z})^3 \ll \Lbox^3$.}, the time wasted in such a way can be orders of magnitude larger than the one employed to integrate the motion of particles within the lightcone. A way to partially compensate for this is to employ the same simulation to produce multiple lightcones, \ie placing multiple observers in the simulation box. However, such approach entails that the lightcones produced are not independent, as they sample the same region of the universe and thus the same structures (although at different times in their evolution). In the following, we focus on simulations that aim at producing a single full-sky lightcone for an observer placed at the center of the simulation box. The reason for this choice is that forthcoming galaxy survey will cover a fraction $O(1)$ of the sky.

The concept described above is exemplified in Fig.~\ref{fig:scheme_lc} where we plot a $1+1$-dimensional projection of the lightcone evolution (solid lines) as a function of lookback time, converging at
the observer location at the present time while diverging at the Big Bang, respectively located at the bottom and the top of the plot. 
In this example, we highlight a portion of Universe with linear size $\Lbox \approx 7000 \, \hMpc$ by enclosing it in vertical dot-dashed lines. When such a region is simulated (visually represented with an evolving cosmological density field, taken for display purposes only from a simulation with box size $\Lbox=512 \, \hMpc$ ), the particles' equations of motion are integrated between discrete timesteps $\mathrm{d}t$ (horizontal grey lines). 
Particles are saved in the lightcone output when they encounter the solid lines (\ie the lightcone boundary) during their time integration (which progresses from top to bottom in the Figure).

All particles below the $\pm \Rlc$ lines (\ie within the gray-shaded area) are therefore simulated although never used (again) for the lightcone construction. Even assuming the computational time is evenly distributed across cosmic time (while in reality matter clustering increases the computational cost at low redshift), it can be seen that a large fraction of the computing time is spent on particles that are then discarded. The Figure also exemplifies how, in order to obtain a lightcone out to some target redshift (horizontal dashed line) without tiling simulation replicas, the box size required (vertical dot-dashed lines) quickly increases.

As we mentioned in the previous section, in a fully-relativistic setup, particles outside the lightcone do not affect the evolution of those inside, and hence their integration can be stopped as soon as they leave it with no consequence on the result (as it is done in the SDF). However, the vast majority of cosmological codes work in a Newtonian approximation, 
computing the gravitational field assuming an infinite speed of light (\ie the gravitational field at any given time depends on the full matter distribution at such time). 
Therefore, simply stopping the integration of (or even removing) particles outside of the lightcone is not a suitable approach.

In order to overcome this limitation, the DZS algorithm combines the relativistic notion of lightcone with the Newtonian nature of typical cosmological simulations, in a way which is fully consistent with the latter and does not require the gravitational solver to be changed. 
Hence, it can be integrated with minimal modifications in most existing N-body codes. Such property of the DZS method is particularly valuable, as an array of auxiliary features, physical and processing modules, and optimizations already implemented in the baseline codes can be used alongside DZS, and benefit from the improved computational performance. In order to test the fidelity and speed-up obtained using the DZS approach, we have implemented it in the popular \gadget code, which is briefly reviewed in the following Section.

\subsection{The \gadget code}
\label{sec:gadget}
In the following, we briefly summarize the main features of the code \gadget, with emphasis on the ones that are most relevant for our implementation of the DZS algorithm. We refer the reader to \citet{gadget2} for a more-detailed description.

\gadget follows the evolution of dark matter (DM) and baryons, discretizing them into particles. It employs a Smoothed Particle Hydrodynamic (SPH) approach to solve the hydrodynamics equation, while 
an hybrid TreePM algorithm is used to compute gravitational forces. In this work, we focus on DM-only simulations, where SPH is not relevant, and hence we only describe the features relevant to gravitational interactions. However, the DZS algorithm can be easily extended to any additional implementation employing particles as tracers of some underlying physical quantities.

The hybrid algorithm employed by \gadget combines the speed, long-range accuracy and memory efficiency of the particle-mesh algorithm \citep[PM, see \eg][]{Klypin+1983, White+1983} with the high dynamical range and short-range precision of a tree algorithm, in the so-called TreePM algorithm \citep{Xu1995,Bode+2000,Bagla2002,Bagla+2003}. The former uses an auxiliary (uniform) grid to compute the matter density and determines the gravitational force by solving the Poisson equation in Fourier space. 
In the tree algorithm, on the other hand, 
all the particles in the simulation domain populate an oct-tree, and the gravitational acceleration exerted on a given particle by all the other ones is computed through the multipole expansion of the gravitational attraction of each tree node. In \gadget this expansion stops at the monopole order. The size of the tree nodes employed (\ie how `deep' the oct-tree is traversed) is determined dynamically employing one of two available tree-opening criteria: a geometrical one comparing the node linear size with its distance from the particle considered, and a relative one limiting the (estimated) force error introduced by the multipoles expansion. 
The PM and tree-based gravitational force estimations are then combined in Fourier space using an exponential kernel, ensuring a smooth transition between the tree-dominated (inner) region and the PM-dominated (outer) one.

The \gadget tree structure is implemented as a linked list of nodes. Each node represents a portion of the simulation box and stores both physical properties of the particles (\eg total mass, center-of-mass position and velocity) in such region of space and a series of links used to traverse the tree. The resulting $3$-dimensional oct-tree represents a hierarchical structure of nodes: the top --~or root~-- node coincides with the simulation domain, while the tree level below it is made of $8$ child nodes, each of which covers one octant of the parent node. This partition is recursively repeated at each level of the tree for every node containing more than one particle. We adopt a common terminology to indicate the relationship between such nodes, using the terms parent, child, and sibling node to indicate the ones in the upper, lower and same tree level of the considered node, respectively. Finally, \gadget is a massively-parallel code. The computation, particle data, and oct-tree are distributed over multiple computing tasks using the Message Passing Interface \citep[MPI,][]{MPI}. 
The connection between tree branches residing on different tasks is provided by pseudoparticles, \ie particle-like structures that store the center-of-mass properties of each tree \textit{node} stored on another task.

In order to cope with the large dynamical range typically encountered in astrophysical systems, \gadget allows particles to have individual timesteps, determined by their dynamical properties. For instance, high-density environments can have orders-of-magnitude smaller timescales than the intergalactic medium.
In particular, particle timesteps in \gadget are organized in power-of-two subdivisions of the system timestep $\mathrm{d}t$, \ie individual particle timesteps are rounded to the closest smaller interval $\mathrm{d}t_n \equiv \mathrm{d}t / 2^n$. Hence, all particle timesteps are commensurable to each other and, therefore, it is ensured that there exist simulation steps --~called \textit{global}~-- in which all particles are actively evolved.

Finally, in \gadget, particles can belong to six different groups --~or \textit{types}~-- each one of them potentially having unique properties, most notably their mass and gravitational softening length $h$. Typically, different particle types are used to represent either (astro)physically-different (\eg gas, stars, black holes, etc.) or numerically-different (\eg particles with different resolution) objects, as well as a mix of the two. We anticipate here that we will indeed employ different types for the latter.

\subsection{The DZS algorithm}
\label{sec:algorithm}
Our implementation of the DZS algorithm builds on the oct-tree structure of \gadget 
as an independent module which is called recursively during the simulation. When the DZS module is invoked, the entire simulation oct-tree is `walked' starting from the root node. 
Whenever a node does fulfill a set of node-merging criteria, its children nodes 
are replaced by a single particle inheriting all their physical and numerical properties, as detailed in the next paragraphs. If the criteria are not satisfied, the process is repeated separately for each child node. This approach requires only a single tree walk (which scales with the logarithm of the number of particles), making it very efficient. In all our tests, the time required by the execution of the DZS module itself is negligible, never exceeding 0.1\% of the entire simulation time. This does not include any additional overhead that may arise from the work-load imbalance due to the (potentially large) difference in resolution across the simulation domain, which will be discussed in detail in Sec.~\ref{sec:imbalance}.

The error introduced by the suppression of the simulation resolution in the nodes that fulfill the merging criteria (corresponding to regions of the simulation domain lying outside the observable lightcone, as will be detailed below) can be arbitrarily reduced by tweaking the node-merging criteria themselves, at the cost of a smaller performance gain. We have implemented the node-merging criteria to resemble the ones used in the force computation (see Sec.~\ref{sec:refinement} for more details). Hence, if the parameters used 
are equal, the error introduced by DZS at any given timestep and for any given particle inside the lightcone is ensured to be \textit{at most} the one introduced by the force computation. Typically, however, we advise to employ more stringent criteria for DZS, as its effects are permanent and generally non-isotropic for any given particle (unlike, in principle, the error introduced by the force computation). 

\begin{figure*}
    \centering
    \includegraphics[width=0.75\textwidth]{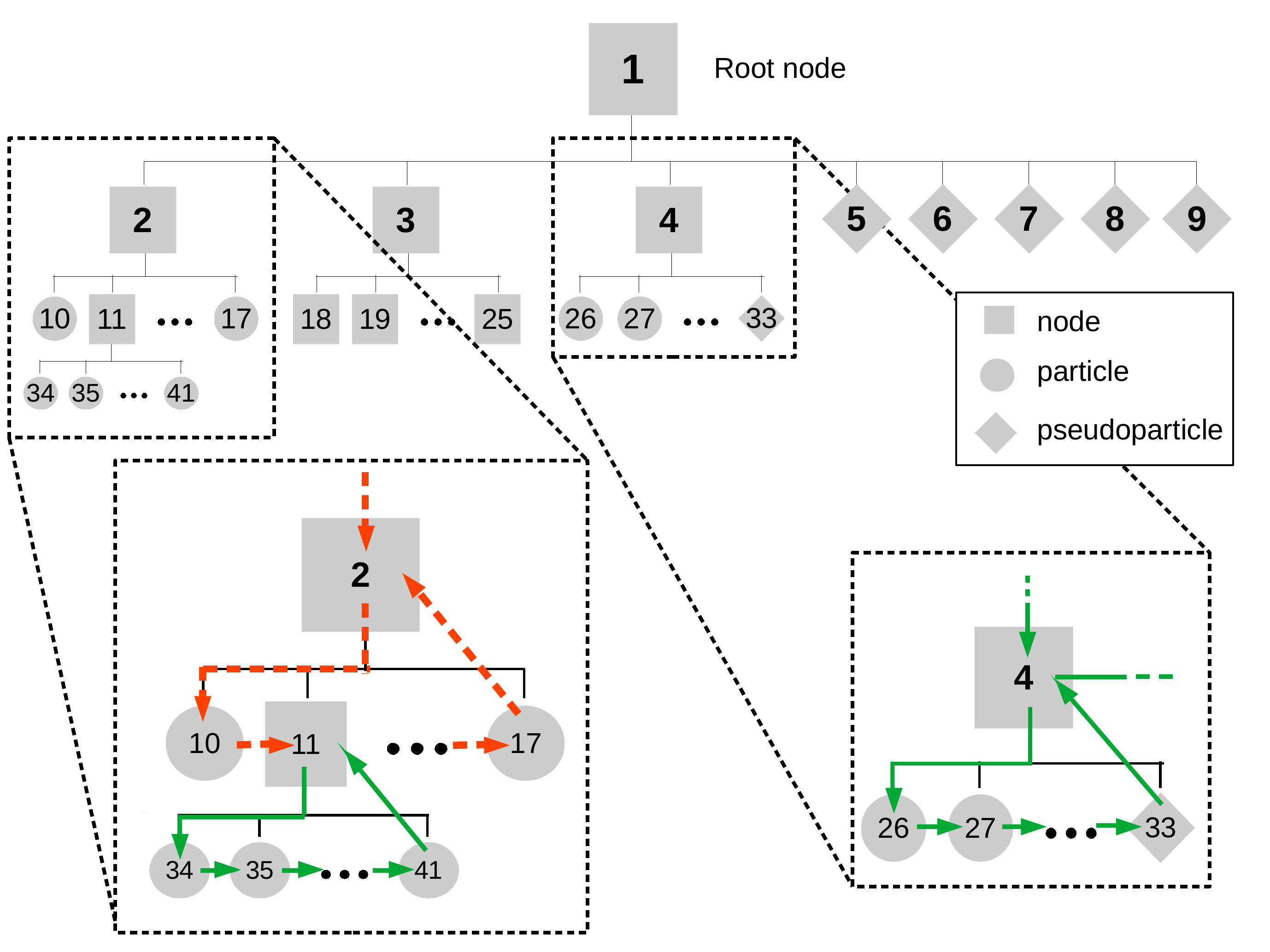}
    \caption{Flowchart of the tree walk in the DZS algorithm. The two boxes in the lower part of the Figure show two examples of tree walk, using a simple color scheme with green solid lines signaling that particles encountered during the walk should be de-refined, and red dashed lines signifying that they should not be merged.}
    \label{fig:treewalk}
\end{figure*}

In Fig. \ref{fig:treewalk} we provide two examples of how the node-merging procedure of the DZS algorithm works in practice. The Figure depicts the tree structure on a single MPI task hierarchically distributed along the vertical direction, where each grey box represents a node, the thin black lines show its connections to other nodes, particles are depicted by grey circles, and diamond symbols are used for pseudo-particles. The two boxes show two details of the node-merging process, using a simple color coding where red dashed lines signal that particles encountered during the walk should not be merged, while green solid lines flag that particles should be de-refined. In the lower left box, the tree walk (red dashed line) reaches node $2$, the merging condition is checked but not met. The tree walk continues from the first child node ($10$) and, since the latter is a particle, the walk continues from its sibling $11$. In this case, the merging criterion is met and, therefore, the node is flagged for merging (depicted by the tree-walk line becoming solid green), and the walk continues from its first child ($34$). The latter is a particle, and since its parent node satisfies the merging condition, it is flagged for de-refinement before the walk continues from its sibling ($35$). When the last child node is processed, the walk returns to the parent node ($11$).

A slightly different configuration is presented in the bottom right box. In this case, one of the child nodes is a pseudoparticle ($33$). 
Since the tree is consistent across different tasks, 
the de-refinement process 
produces exactly the same outcome independently of the particle distribution among tasks. Hence, pseudoparticles are safely ignored, as the task containing the data pointed to by such pseudoparticles will process them and is guaranteed to accomplish identical results.

Once the full tree has been walked, particles flagged to be de-refined are removed and tree nodes flagged to be merged are turned into particles. This operation is done by transferring the center-of-mass properties of the flagged node to the newly-created particle.\footnote{This is sufficient in our case since \gadget stores enough information in the tree nodes to generate a fully-functional new dark matter particle. In general, the same outcome can be achieved by explicitly computing average properties of the de-refined particles, at the price of additional communication in the case a node contains pseudo-particles.} If nodes flagged for de-refinement only contained particles local to the task, the new low-resolution particle is simply created \textit{in loco}; however, whenever a flagged node contains a pseudoparticle --~i.e. the tree branch is partly contained in another task~-- the newly-created particle will reside on the task whose rank is the minimum among all tasks 
contributing to 
that tree branch. This precautionary measure ensures that only a single new particle is created for branches mapped to several tasks and does not require 
communication, which could be a source of performance loss.

As mentioned, \gadget allows particles to have individual timesteps. 
To maintain the time-integration consistency, particles grouped together by DZS need to be synchronized. This is ensured by applying the DZS algorithm only during so-called global timesteps.

By removing and replacing particles within the simulation, DZS drastically changes the particle 
distribution across tasks, as well as the tree structure. For this reason, whenever DZS modifies at least one particle, the tree needs to be rebuilt, the acceleration and timestep of new particles recalculated, and a new domain decomposition performed. To mitigate (and in fact practically remove) the overhead introduced by these operations, we additionally restrict DZS to be invoked only whenever a new domain decomposition (which is the most time-consuming of them) is performed. By running DZS just before a standard domain decomposition, we ensure that the computational effort invested into it is not duplicated. Finally, we force individual particle timesteps outside of the lightcone to be \textit{at most} as short as the shortest timestep inside it. While most of the times this condition has no effect, it can prevent some pathological situation where two particles outside of the lightcone get unphysically close as a consequence of the de-refinement procedure, and hence require a very short timestep to precisely integrate their dynamics, despite the latter has a negligible impact on the evolution of particles inside the lightcone.

By grouping particles together, DZS changes the mass of individual resolution elements in the simulation while preserving the total mass in the system, unlike SDF. Although taking place outside of the lightcone --~where the detailed dynamics of particles is of no interest~-- the evolving particle mass can potentially alter the large-scale density field and increase the computational cost of the simulation. The latter is a consequence of the larger gravitational pull of de-refined particles (due to their increased mass) that can significantly decrease the timestep of nearby particles if not compensated by an appropriate scaling of their gravitational softening. 
Keeping in mind that the 
latter 
is related to the ideal volume occupied by the density element represented by each zero-dimensional particle, DZS rescales the original softening length by a factor $(M/M_\mathrm{init})^{1/3}$, where $M_\mathrm{init}$ is the initial (\ie before de-refinement) mass of particles.

The merging criterion employed by DZS is designed to ensure a precise integration of particles inside the lightcone (see Sec.~\ref{sec:refinement}). However, the evolution of the large-scale gravitational field needs to be approximately followed as well, since it affects the particles inside the lightcone in the Newtonian approximation. For this reason, the merging criterion is complemented with the possibility to limit the maximum linear size $\lmax$ of the nodes flagged to be merged. This is similar to setting the background resolution in zoom-in simulations. 
Notice that even the very conservative choice of $\lmax = 2 l_\mathrm{min}$, where $l_\mathrm{min} = \Lbox / N_\mathrm{p}^{1/3}$ is the smallest node size if particles were homogeneously distributed, already significantly speeds up the computation approximately by a factor $\gtrsim 16$ \textit{outside} of the lightcone\footnote{The factor $\gtrsim 16$ is made up by a factor $8$ coming from the decreased number of particles, a factor of $2$ coming from the increased mean inter-particle distance that allows larger timesteps, and an additional gain from the reduced clustering, since the merging criterion is geometical and, hence, clustered particles are merged into single de-refined particles.} 

Finally, our implementation of DZS allows de-refined particles to be moved to a different particle type with respect to the full-resolution particles, to allow for an easier handling of the simulation dataset for both on-the-fly and post-processing operations.

\subsubsection{De-refinement criteria}
\label{sec:refinement}

\begin{figure}
    \centering
    \includegraphics[width=\columnwidth]{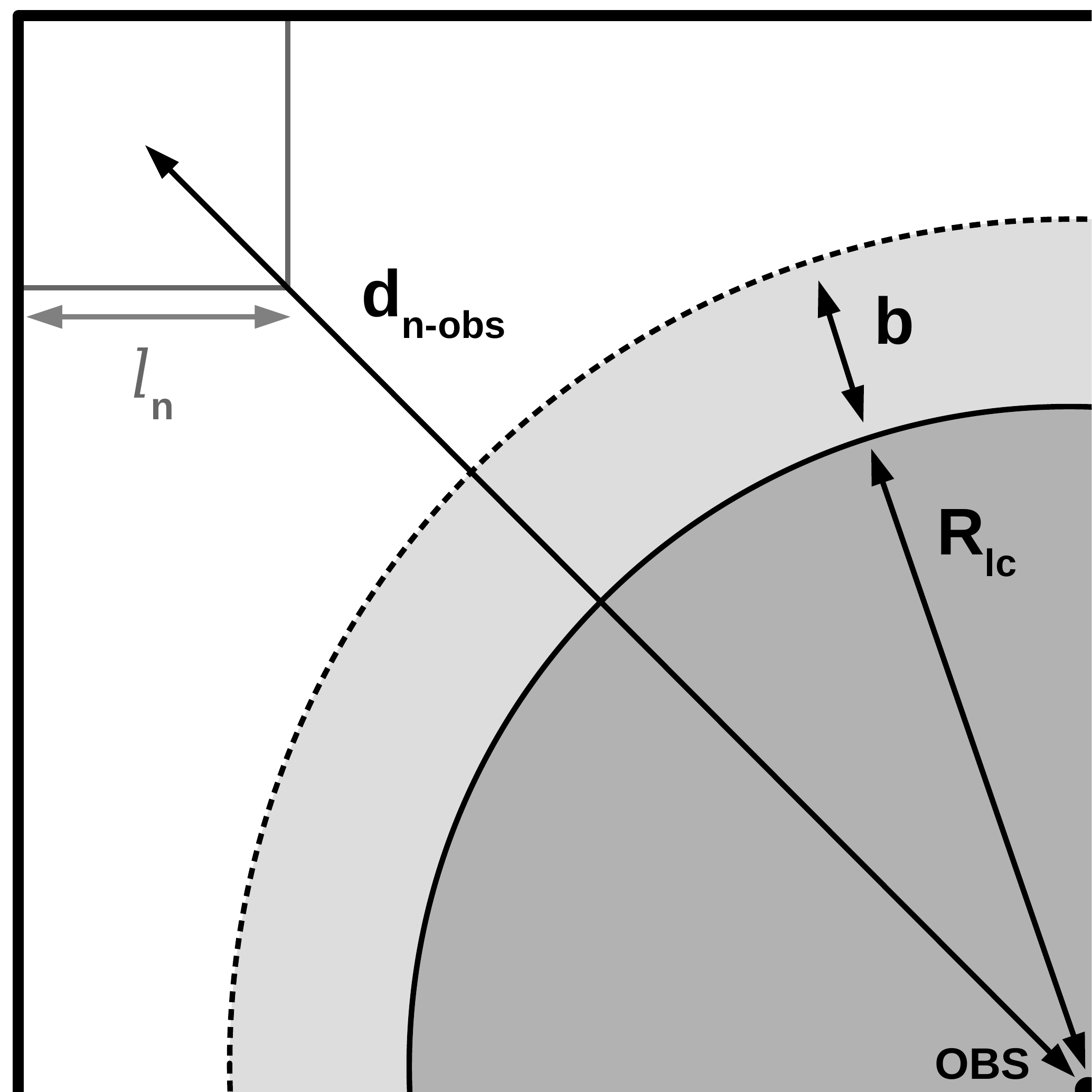}
    \caption{Schematic representation of the relevant lengths involved in the de-refinement process. We are showing here a 2D projection of one octant of the simulation box. The observer is located at the bottom right corner (corresponding to the center of the box, while the solid circle and dark-gray shaded area represent the lightcone radius and volume, respectively. The light-gray shaded region (delimited on the outside by the dashed circle) denotes the buffer zone. Finally, a single tree node is depicted in the top left of the plot.}
    \label{fig:scheme_node}
\end{figure}

Since the de-refinement criteria of the DZS scheme are based on the tree-opening criteria used in the force calculation in \gadget, we first review the latter before introducing their adaptation to DZS. The relevant quantities and geometry are schematically depicted in Fig.~\ref{fig:scheme_node}.

\gadget allows for two different opening criteria in the force calculation, one geometrical and one dynamical. The former requires that, for each target particle $i$ for which the force is to be computed, a given node $n$ must be opened (i.e. the tree walk should be continued through its child nodes) whenever its linear size $l_n$ is larger than the distance between its centre of mass and the particle $d_{in} = |\mathbf{r}_i - \mathbf{r}_n|$ times an (user-defined) `opening angle' $\tgeom$, \ie 
\begin{equation}
\frac{l_n} {d_{in}} \geq \tgeom .
\end{equation}

The second --~dynamical~-- criterion is devised to strictly enforce a maximum force error introduced by the tree-opening choice and, at the same time, to allow large nodes to be used whenever possible. It takes the form
\begin{equation}
\label{eq:derefinement_dynamical}
   \frac{G M_n}{d_{in}^2} \left( \frac{l_n}{d_{in}} \right)^2 \geq \adyn \ |\mathbf{a}_{i,\mathrm{old}}|
\end{equation}
where $G$ is the gravitational constant, $M_n$ is the total mass contained in the node, $\mathbf{a}_{i,\mathrm{old}}$ is the particle acceleration in the previous timestep, and $\adyn$ is a (user-defined) parameter that sets the force accuracy. 

These opening criteria are local, since they are used -- and yield different results -- for each particle $i$ in the simulation. However, they need to be cast into global criteria to be usable by DZS, since its modifications to the simulation affect all particles. In addition, the force computation is `particle-centered', while DZS requires `node-centered' criteria. To achieve this, we have replaced the distance $d_{in}$ 
with the minimum distance of the node $n$ to any given particle lying inside the lightcone, allowing for an additional spherical shell of user-defined width $b$ that acts as a buffer zone. In practice, $d$ is replaced by 
\begin{equation}
    \mathcal{D}_n = \min_{i:\ |\mathbf{r}_i -\mathbf{r}_{obs}| < \Rlc+b} |\mathbf{r}_n - \mathbf{r}_i| = \max \left(  d_{n-obs} - \Rlc - b , 0 \right)
\end{equation}
where $d_{n-obs}$ denotes the distance between the center of mass of the node and the center of the lightcone (\ie the observer's location $\mathbf{r}_{obs}$). 

While this replacement is sufficient to make the geometric criterion global, the dynamical one depends on an additional local quantity, namely the last-recorded acceleration of the particles. For this reason, we have replaced $a_{i,\mathrm{old}}$ with its minimum value among all particles inside the lightcone.

Finally, we note here that the buffer zone can be used in order to enforce a minimum volume (\ie a sphere of radius equal to the buffer length $b$) to be resolved at the original resolution down to $z=0$. This allows (i) the production of a traditional time-slice output of a desired fraction of the simulation box; and (ii) to produce multiple lightcones (for observers at different locations in the box) within the same run, which are independent from each other out to some redshift $z$ (the exact value depends on the observer locations).

In the remainder of the paper, unless otherwise specified, we will adopt a standard configuration of DZS employing a geometrical opening criterion with angle $\tgeom = 0.1$, a maximum node length of $\lmax = 4 \lambda$, where $\lambda$ is the mean inter-particle distance (\eg ensuring the low-resolution region is sampled with an equivalent resolution of \textit{at least} $128^3$ for a simulation with $512^3$ particles), and a buffer size corresponding to $5 \lambda$. These are relatively conservative parameters. The effect of different parameter choices are investigated in Sec. \ref{sec:parameter}.

\subsubsection{Future developments}
Before moving on to a thorough validation of the DZS algorithm in the following Section, we want to briefly mention here some foreseeable developments that can --~or are going to~-- be included in DZS. The structure of DZS is such that it can be easily extended to all particle-based implementations of additional physics that may be already available in the code, or that may be included in the future. For instance, \gadget is already capable of simulating a wide variety of physical processes beyond the (nowadays standard) adiabatic and radiative hydrodynamics, as \eg Modified Gravity \citep[\eg][]{MG-gadget}, non-standard Dark Sector physics \citep[\eg][]{C-gadget,Codecs}, massive neutrinos \citep[\eg][]{Viel+2010}, Ultra-Light Axions \citep[also known as Fuzzy Dark Matter, see \eg][]{Ax-gadget}, radiation-transport \citep[\eg][]{TRAPHIC, Petkova&Springel2009}, magnetic fields \citep[\eg][]{Dolag+Stasyszyn2009, Bonafede+2011}, and many more. As all these implementations are still resorting on a particle-based discretization of physical processes, such additional modules can be combined with the performance improvement granted by DZS in a relatively-easy way. In particular, we are working on extending our current implementation of DZS to include modified gravity (Nori et al., in prep).

Furthermore, the lightcone output procedure that is currently paired with the DZS algorithm can be easily --~and independently from the latter~-- extended to include a range of additional properties, computed on-the-fly during the simulation run, in order to compensate for the lack of snapshot outputs. In particular, we plan to extend the current lightcone output to include the gravitational potential and its time derivative.

Finally, a possible further optimization (which however is independent from DZS itself) consists in employing initial conditions with coarser resolution outside of the region of space which ends up in the lightcone at the redshift of interest. This allows to include large-scale density modes with a limited additional burden for the simulation.

\section{Algorithm validation}
\label{sec:tests}

We now present the results of a test suite designed to validate the accuracy of DZS. These simulations are conceived --~for comparison purposes~-- to mimic and extend the ones performed by L17. They cover a factor of $16$ in box size and a factor of $8$ in number of particles. We complement them with a re-run of the $\Lambda $CDM simulation of the {\small DUSTGRAIN} suite (a 2 Gpc/$h$ cosmological box with $2048^{3}$ CDM particles, Baldi et al., in prep.), in order to test our algorithm in a more realistic and computationally challenging setup, and to extend the range covered in particle number $N_\mathrm{part}$ by another factor $8$. The main properties of our test suite of simulations are summarized in Table~\ref{table:test_L17}.

In the following we will use the term `twin simulations' to indicate pairs of runs that only differ in the usage of DZS, which is enabled in the one labeled as \textit{dzs} and disabled in the one labelled as \textit{std}.

\begin{table}
    \centering
        \begin{tabular}{lccc}
        \hline
        Name & $\Lbox \, (\hMpc)$ & $N_\mathrm{p}$ & Modeled on\\
        \hline
        \textsc{tiny}$^a$    &  $128$ &  $512^3$ & \\
        \textsc{small}   &  $512$ &  $512^3$ & \citet{Llinares17}\\
        \textsc{medium}  & $2048$ &  $512^3$ & \citet{Llinares17}\\
        \textsc{large}   & $8192$ &  $512^3$ & \citet{Llinares17}\\
        \textsc{largeHR} & $8192$ & $1024^3$ & \citet{Llinares17}\\
        \textsc{DG}      & $2000$ & $2048^3$ & Baldi et al. (in prep)\\
        \hline
        \multicolumn{4}{l}{$^a$ Used only in performance estimation (Sec.~\ref{sec:estimation}).}
        \end{tabular}
    
    \caption{Simulation set employed for scaling and accuracy tests, replicating the test suite of L17. Each row corresponds to two twin simulations, one run using the DZS algorithm and one without.}
    \label{table:test_L17}
\end{table}

\begin{figure*}
    \centering
    \includegraphics[width=0.95\textwidth]{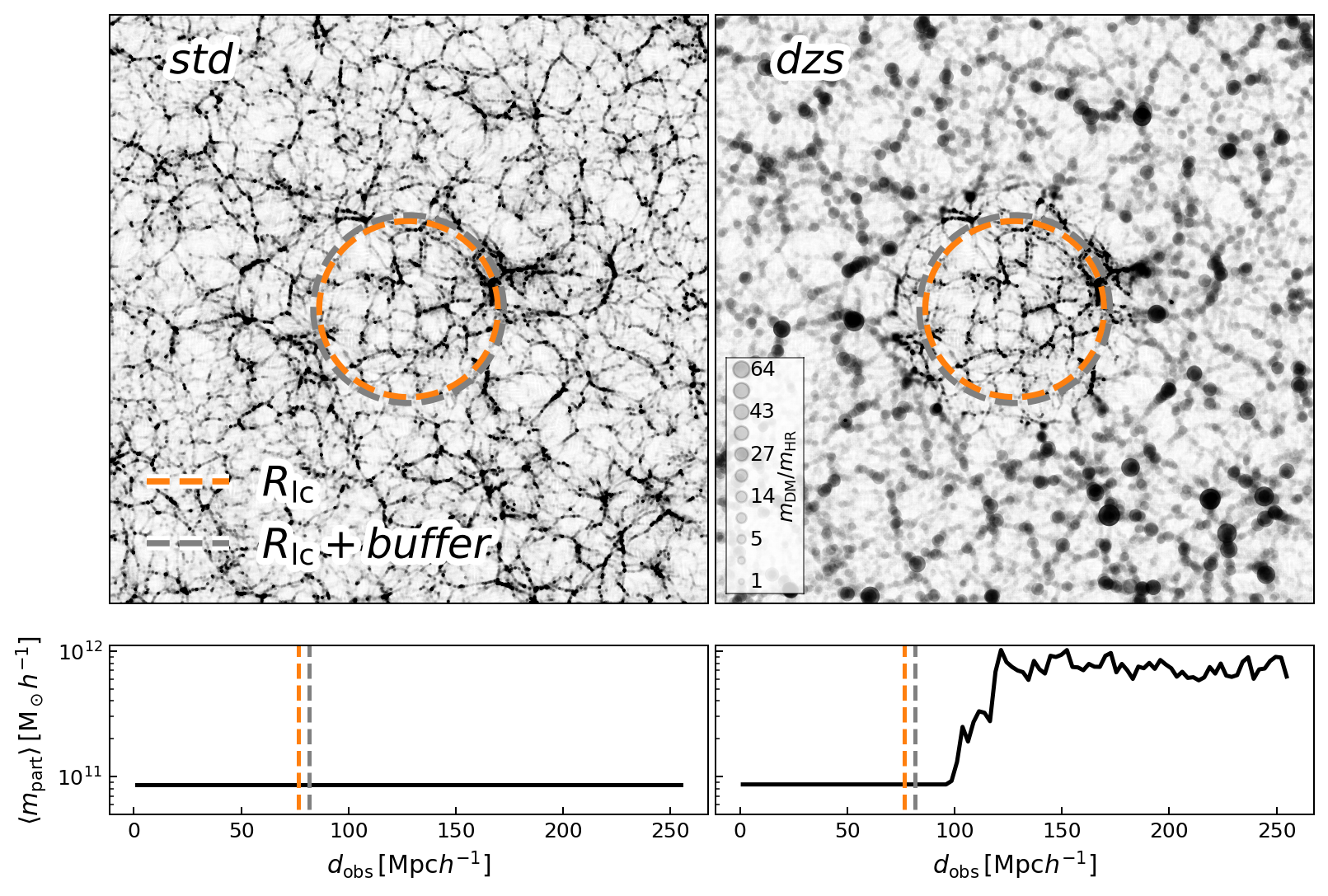}
    \caption{Top panels: Dark matter distribution in a slice of thickness $0.5\,\hMpc$ extracted from the \textsc{small} twin simulations, at $z=0.05$. Each dot represents a particle in the simulation, with size and opacity increasing with their mass (as indicated in the bottom left corner of the right panel). In the \std run (left panel), each particle has the same mass, while in the \dzs one, different levels of de-refinement can be seen. The structures inside the lightcone (orange dashed line) are identical in the two runs, and the large-scale matter distribution is preserved also outside of the lightcone in both panels. The grey dashed line shows the effect of adding a buffer zone around $\Rlc$. Bottom panels: the mean particle mass as a function of distance $d_\mathrm{obs}$ from the observer (placed at the center of the box). }
    \label{fig:showcase}
\end{figure*}

In Fig.~\ref{fig:showcase} we provide a view of the effect of DZS on the simulated particles in the \textsc{small} twin runs. Each top panel shows a slice of thickness $0.5\,\hMpc$ at redshift $z=0.05$ extracted from the \std run (left) and from the \dzs one (right). Each particle in the slice is plotted as a black dot, with size and opacity increasing with the particle mass (linearly and quadratically, respectively, for the sake of visual clarity, as indicated in the bottom left corner of the top right panel). It can be clearly seen that the structures inside the lightcone (dashed orange circle) are unaffected by DZS, while outside of it the number of particles in the right panel rapidly decreases, but the large-scale structures are preserved. 
We emphasize this feature by showing in the bottom panels the average particle mass as a function of distance from the observer (placed at the center of the box).
The latter does not increase immediately outside of the lightcone (plus buffer) radius since a geometrical de-refinement criterion is used in the \textsc{
small} \dzs run, and hence only nodes with angular size (as viewed from the closest point on the buffer surface) below a given threshold are merged together. Finally, we note that the noise at large radii is due to the spatial nature of the oct-tree subdivisions. When a low-resolution particle is produced, its mass will reflect the mass contained in the parent node, and hence the mass in the region of the universe covered by the latter. It follows that nodes encompassing large cosmic structures (\eg galaxy clusters) will contain a larger mass than those enclosing voids, producing a disparity in the mass of low-resolution particles stemming from them, which ultimately leads to the noisy profile in the Figure.

\subsection{Lightcone observables}
\label{sec:observables}
In the following we move to a quantitative evaluation of the accuracy of (our implementation of) DZS. 
The different simulation paradigm employed by DZS renders typical measures of accuracy (\eg the matter power spectrum, the halo mass function, etc.) not readily usable, and adapting them to the new framework would produce issues when comparing them to other results. Hence, we decided to base our accuracy analyses only on observable quantities, \ie measured (using a piecewise-constant approximation) on the 4D-lightcone surface. The latter corresponds to the 3D volume generated as the set of spherical surfaces centered on the observer with radius $\Rlc (z)$ for any redshift $z \leq z_\mathrm{max}$, with $z_\mathrm{max}$ being the highest redshift considered for a given lightcone.
This choice is also the most conservative in terms of accuracy testing, since by definition the particles on the lightcone spatial boundary are those most affected by the decreased resolution outside of it, being the closest to the low-resolution regions. Additionally, in Sec.~\ref{sec:accuracy} we investigate the time-integration fidelity of DZS by comparing time slices in twin simulations. Finally, it is important to stress again here that the DZS algorithm can be made arbitrarily precise by tweaking the node-merging criterion parameter, at the cost of a decreased speedup (see Sec.~\ref{sec:parameter} for details). Hence, we test the accuracy only in our standard set of settings, also in order to be consistent with the performance tests presented in Sec.~\ref{sec:performance}.

\subsubsection{Halo Mass Function}
\label{sec:hmf}

We start our investigation by comparing the lightcone halo mass function (LCHMF) integrated over the redshift range $0.0 \le z \le 0.68$ in the \textsc{DG} twin simulations. Among our simulation pairs, this is the one that better resolves structures (\ie it has the best mass resolution) and, simultaneously, has a box size large enough to (i) allow the construction of a lightcone out to a redshift significantly larger than zero ($z \sim 0.68$) without requiring box replications, and (ii) is affected by DZS for a significant fraction of the simulation time. In comparison, the \textsc{largeHR} pair is affected by DZS for a much longer time but has a $550$ times worse mass resolution, hindering the identification of many cosmic structures.
In order to produce the LCHMF we save with high cadence the halo catalogs produced by the Friend-of-friends (FoF) algorithm integrated in \gadget. These outputs were then combined into a lightcone catalog by means of a post-processing procedure employing a piecewise-constant approximation, ensuring that this construction only makes use of haloes entirely inside the lightcone at the time of the output.

\begin{figure}
    \centering
    \includegraphics[width=\columnwidth]{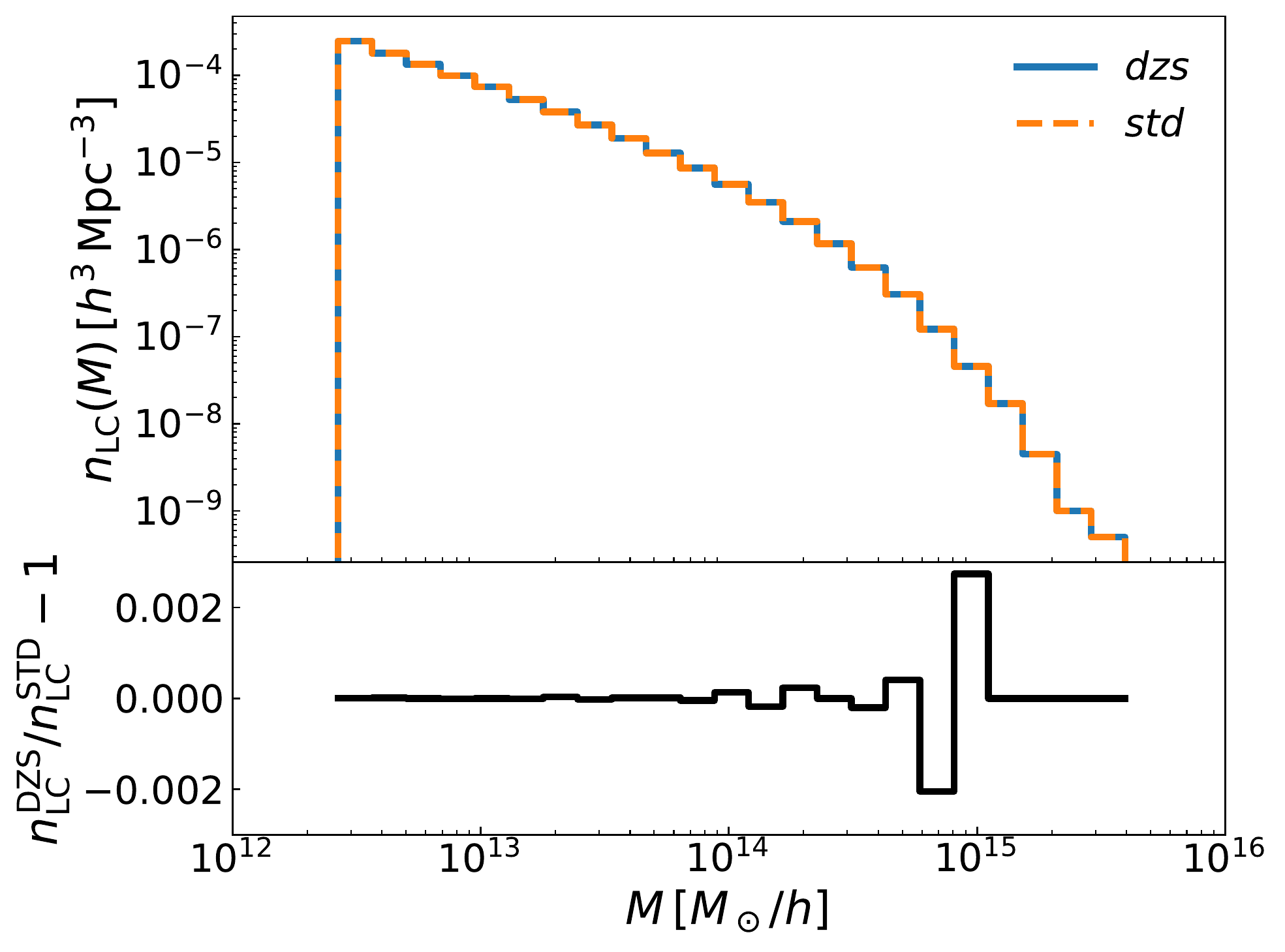}
    \caption{Lightcone halo mass function (top panel) at redshift $0.0 \le z \le 0.68$ in the \textsc{DG} simulations with DZS enabled (solid histogram) and disabled (dashed histogram). Their relative deviation is always below $0.2$\% (bottom panel).}
    \label{fig:hmf}
\end{figure}

The results of this procedure are shown in the top panel of Fig.~\ref{fig:hmf}, where we show the redshift-integrated mass spectrum of haloes on the lightcone. It is immediately clear that DZS is able to almost-perfectly recover the LCHMF of the standard simulation. More quantitatively, as shown in the bottom panel of the Figure where we plot the fractional deviation of the LCHMF computed in the \dzs run with respect to the \std one, our algorithm can reproduce the LCHMF with a sub-percent level accuracy.

\subsubsection{Sky-projected lightcone}
While the LCHMF is a measure of the accuracy reached by DZS in simulating large cosmic structures, we can alternatively compare the lightcone mass distribution projected on the sky in some redshift interval. In order to obtain such projection, we use HEALPIX \citep[][with parameter $NSIDE=1024$]{healpix} to obtain the original pixelization from the simulation, but then we downsample it to $NSIDE=256$ for display purposes. Although not directly observable, this quantity is the one that large-scale surveys aim to reconstruct starting from either biased tracers (first and foremost galaxies) or weak lensing shear measurements. 

\begin{figure}
    \centering
    \includegraphics[width=\columnwidth]{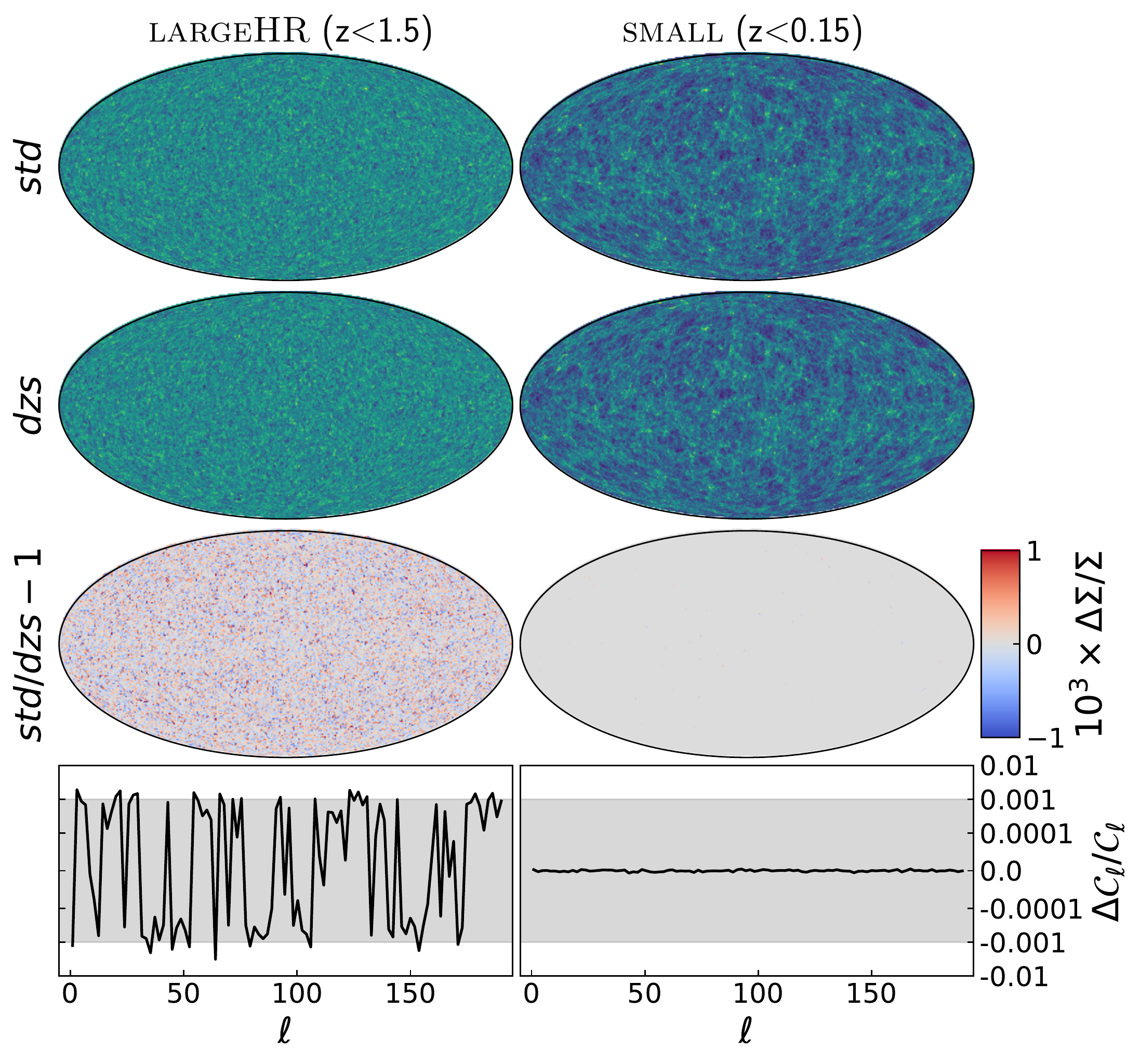}
    \caption{Sky-projected (using HEALPIX sky pixelization with $NSIDE = 256$) density distribution on the lightcone for the \std (top row) and \dzs (second row from the top) runs in the \textsc{largeHR} (left column, integrating over $0 < z < 1.5$) and \textsc{small} (right column, integrating over $0 < z < 0.15$) pairs. The pixel-wise relative difference between the top two rows is shown in the third one. Finally, the last row shows the relative difference in the angular power spectrum computed from the sky maps shown in the top two rows. The shaded gray region indicates the $0.1\%$ accuracy level. Notice that the top two sky maps on the right are identical, and hence their difference is always zero. This means that no particle in the two runs is displaced more than the pixel size. Note that the vertical scale in the bottom panels is linear for absolute values smaller than $10^{-4}$, otherwise the vertical position for a value $v$ is given by $\mathrm{sign}(v) \log(|v|)$.}
    \label{fig:healpix_diff}
\end{figure}

We present such comparison in Fig.~\ref{fig:healpix_diff}. Each column refers to a different simulation pair (reported on the top), and a different redshift range for the lightcone projection (also reported on top). The ranges were chosen in such a way that they encompass most of the time the lightcone has been inside the simulation box. 
The first two rows show the projection in the \std and \dzs runs, while their pixel-wise relative difference is shown in the third row. The latter is very small in most of the pixels, and never exceeds $0.1$\%. Most importantly, the few pixels where it is close to its maximum are randomly located in the sky, showing that they occur because of numerical stochasticity triggered by the slightly perturbed gravitational potential, rather than by systematic effects. Finally, the last row shows the relative difference in the angular power spectrum between the \std and \dzs realisations (i.e. it is not the power spectrum of the difference map). Also in this case, the deviation is almost always below $0.1$\%, indicated by the grey-shaded horizontal band and its scatter (around $0$) appears random.

While Fig.~\ref{fig:healpix_diff} shows the high degree of fidelity achieved employing the DZS approach, it compresses a large redshift range on to the sky. In the left panels in particular, the individual structures are lost and the density field appears random to the eye. For this reason, we show in 
the bottom right panel of Fig.~\ref{fig:3d_lightcone_skymap_combined}
the sky-projected lightcone in the \textsc{DG} \dzs run, but now projected over a thin redshift slice $z \lesssim 0.03$, which allows to appreciate even more how cosmic structures and their renowned web-like topology are unchanged when the DZS approach is employed.

\subsubsection{3D lightcone}
Finally, we investigate here the ability of DZS to capture the full 3D matter density lightcone constructed from our simulations. Unlike the case of the LCHMF, which has been constructed using a piecewise-constant approximation from the halo catalog at fixed cosmic time, we produce the particle lightcone on-the-fly. This is done by saving with very high frequency the particles in a thin spherical shell whose outer rim coincides with the lightcone radius at any given redshift. 

\begin{figure*}
    \centering
    \includegraphics[width=\textwidth]{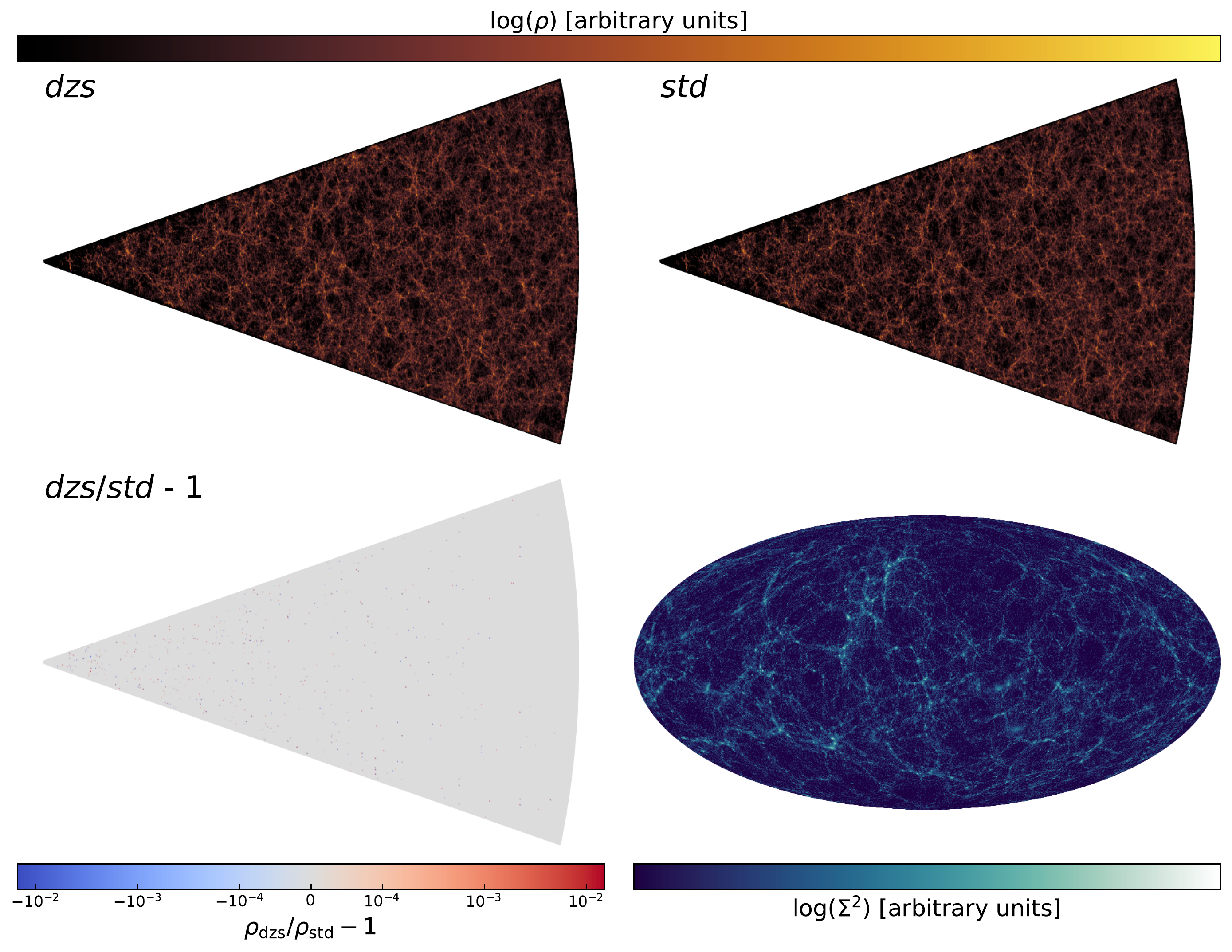}
    \caption{Top panels: Thin slice of the 3D matter density lightcone produced on-the-fly in our \dzs (left) and \std (right) runs of the \textsc{DG} re-simulation. Bottom left panel: Relative difference between the \dzs and \std runs. The color scale is linear for absolute values smaller than $10^{-4}$, otherwise the color for a value $v$ is given by $\mathrm{sign}(v) \log(|v|)$. Bottom right panel:  Sky-projected lightcone (using HEALPIX sky pixelization with $NSIDE = 1024$) density distribution in the \textsc{DG} \dzs run at $z \lesssim 0.03$.}
    \label{fig:3d_lightcone_skymap_combined}
\end{figure*}

In Fig.~\ref{fig:3d_lightcone_skymap_combined}
we show the lightcone constructed as described above in the \dzs (top left) and \std (top right) runs of the \textsc{DG} re-simulation in the redshift range $0.0 \le z \le 0.68$. A visual inspection reveals that the two appear identical. We quantify this in the bottom left panel of the figure, where we show the relative pixel-wise difference of the two runs (notice that this is exactly zero almost everywhere, making it hard to spot pixel-scale differences). It can be appreciated that the perturbations introduced by the DZS algorithm are minor, amounting to only $\sim 0.15\%$ at most (bottom left panel).

\subsection{Integration accuracy}
\label{sec:accuracy}
In the previous Sections, we have shown that DZS is capable of reproducing basic features of the matter distribution, as the LCHMF and the projected matter density across a large redshift range, at a fraction of the cost required by a normal simulation. While these tests show that the DZS algorithm can be faithfully employed, here we test it in the most stringent way possible, \ie comparing \textit{particle by particle} a pair of twin simulations. Note that DZS permanently affects the simulation structure. Hence, the comparisons presented in the following necessarily show the \textit{integrated} effect of DZS over time. For this reason, it is expected that differences will be larger than the nominal force computation error introduced by the node-merging criterion, as the latter is a measure of the error introduced \textit{at each timestep}. By construction, such test makes sense only for particles inside the lightcone. For this comparison, we identify `twin particles' that started exactly identical in the initial conditions 
and 
cross-match them in different runs.\footnote{In the following, we consider the simulation without DZS (labeled \std) as the ground truth. While this could be improved upon by, for instance, employing a direct summation method for the computation of gravitational forces, the usefulness of such comparison would be limited since (i) we are interested in studying the impact of the DZS algorithm only, and therefore prefer to remove any additional potential source of differences between the simulation pairs; (ii) \textit{de facto} all cosmological simulations use approximate methods to compute gravitational interactions (direct summation becomes rapidly prohibitive for large particle numbers and long integration time), and the \gadget code is one of the most widespread in the community. } 

\begin{figure}
    \centering
    \includegraphics[width=\columnwidth]{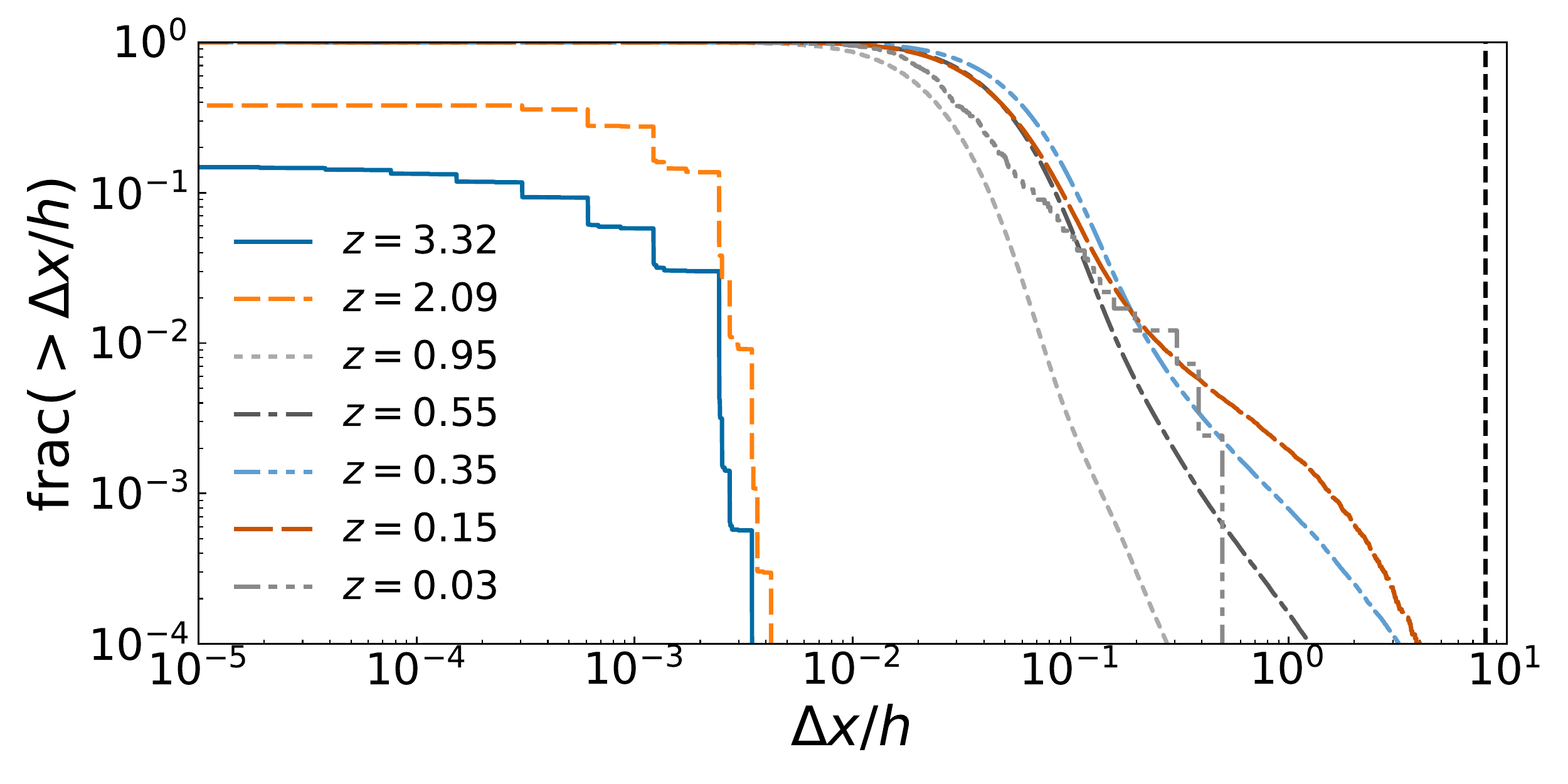}
    \caption{Cumulative distribution of differences in the position (normalized by the softening length) of particles inside the lightcone across the twin simulations \textsc{largeHR}. The vertical dashed line on the very far right indicates $8$ times the softening lengths.}
    \label{fig:displacement_cumul}
\end{figure}

In Fig.~\ref{fig:displacement_cumul} we show the cumulative distribution of displacements $\Delta x$ between particle pairs, normalised by the softening length $h \equiv \Lbox / N_\mathrm{p}^{1/3} / 40$, in the \textsc{largeHR} simulations pair. The vertical dashed line corresponds to $\Delta x / h = 8$, 
a value that we consider a reasonable limit for the spatial reliability of a simulation in terms of individual particle positions. 
In this simulation the first DZS de-refinement occurs at redshift $z \approx 11$. The particle displacement increases with time, as expected, since the small errors introduced by DZS are carried on during the particle integration and therefore pile up. Nevertheless, the displacements always remain very small, with no lightcone particles being displaced by more than $8h$ at any redshift, and only $\approx 0.3\%$ of them being displaced by $h$ at any redshift. Finally, we note that in the lowest-redshift distribution, the number of particles with large $\Delta x / h$ is in fact reduced with respect to the one at $z=0.15$. This is a consequence of the small number of particles remaining in the lightcone (which approaches a vanishing volume at $z=0$) not sampling the tail of the distribution any more.

We have chosen to show the integration accuracy in the \textsc{largeHR} simulation since it is the one where DZS has been active for the longest, and therefore errors have accumulated the most. In fact, the displacement distribution is strongly shifted toward smaller values in the \textsc{medium} and \textsc{small} simulation pairs at any given redshift.

\begin{figure}
    \centering
    \includegraphics[width=\columnwidth]{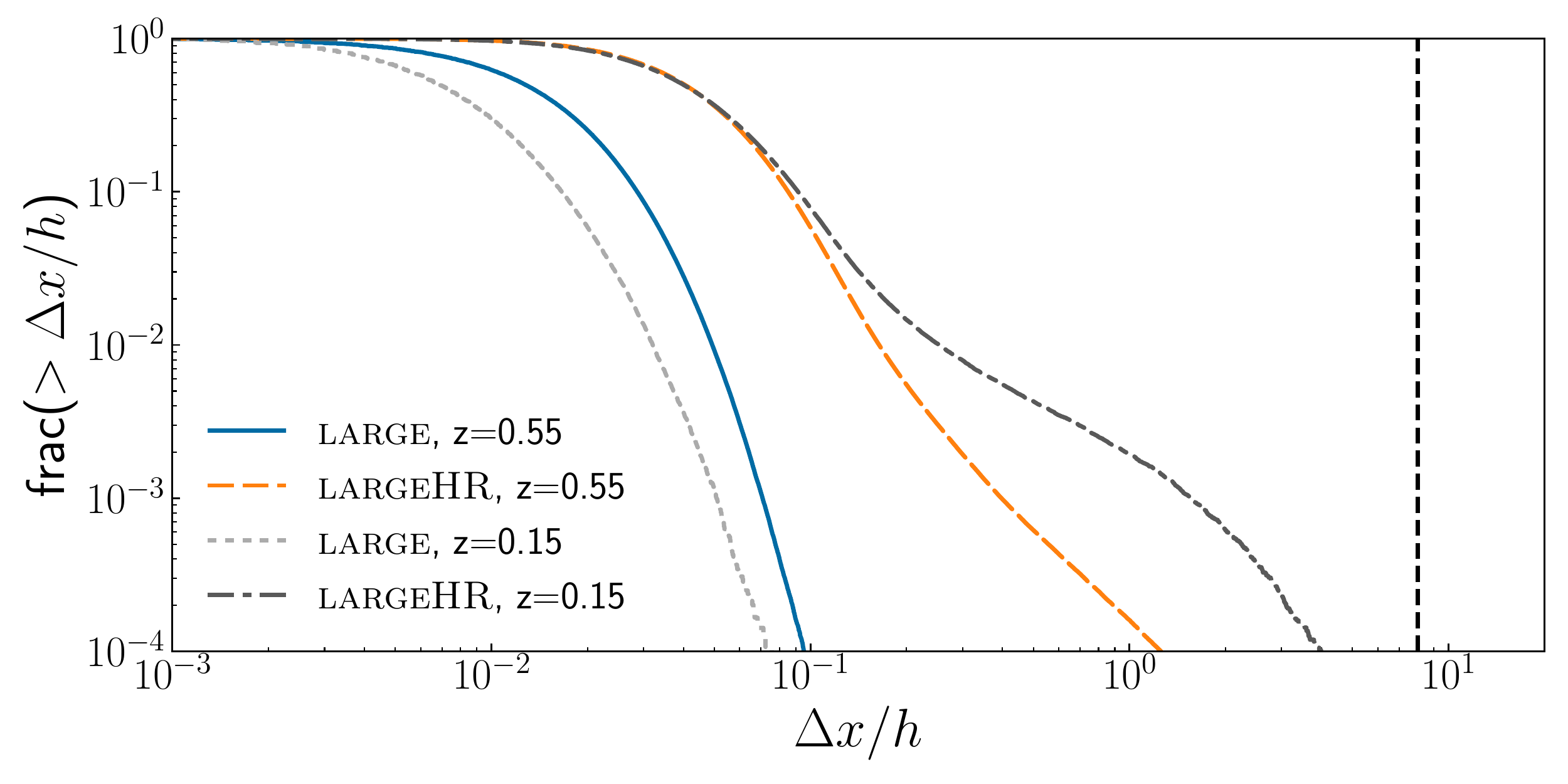}
    \caption{Same as Fig.~\ref{fig:displacement_cumul} but now comparing the \textsc{largeHR} and \textsc{large} pairs, in order to investigate the effect of increased mass resolution. For visual clarity, we show only the curves corresponding to $z=0.55$ and $z=0.15$.}
    \label{fig:displacement_compare_ll81_ll85}
\end{figure}

Finally, in Fig.~\ref{fig:displacement_compare_ll81_ll85} we compare the integration accuracy as a function of resolution. In order to do so, we employ the \textsc{large} and \textsc{largeHR} pairs. 
The increased mass resolution produces two effects: (i) a shift toward larger values of the $\Delta x / h$ distribution, and (ii) the development of a tail containing a small ($10^{-2}$ -- $10^{-3}$) fraction of the particles extending to larger errors. The latter feature (and the distribution itself alike) appears to be very similar to what is found investigating the force errors distribution in \gadget, when a geometrical tree-opening criterion is employed \citep[see \eg the top panel of fig. 1 in][]{gadget2}. In fact, even using a dynamical de-refinement criterion in DZS, we obtain a similar behaviour since, as detailed in Sec.~\ref{sec:refinement}, the necessity for a node-centric criterion renders the dynamical and geometrical ones somewhat similar to each other.

\subsection{Parameter dependence}
\label{sec:parameter}
The comparison presented in the Sections above shows a high level of fidelity in the results obtained employing DZS with respect to those obtained without. However, it does so employing somewhat conservative values for the parameters controlling the behaviour of the DZS algorithm. In a realistic setting, such parameters will be chosen to be as `aggressive' as possible, in order to maximize the performance gain. For this reason, in the following we investigate the dependence of the results on the four different parameters related to the de-refinement process, namely: the geometrical opening angle ($\tgeom$), the dynamical accuracy parameter\footnote{Note that $\tgeom$ and $\adyn$ are mutually exclusive.} ($\adyn$), the buffer size ($b$), and the maximum size a tree node can have and still being merged ($\lmax$). The former two control the transition between the high-resolution zone and the rest, the third enforces a \textit{safety} high-resolution region around the actual lightcone, and the last one sets a minimum resolution for the volume outside of the lightcone.

\begin{figure}
    \centering
    \includegraphics[width=\columnwidth]{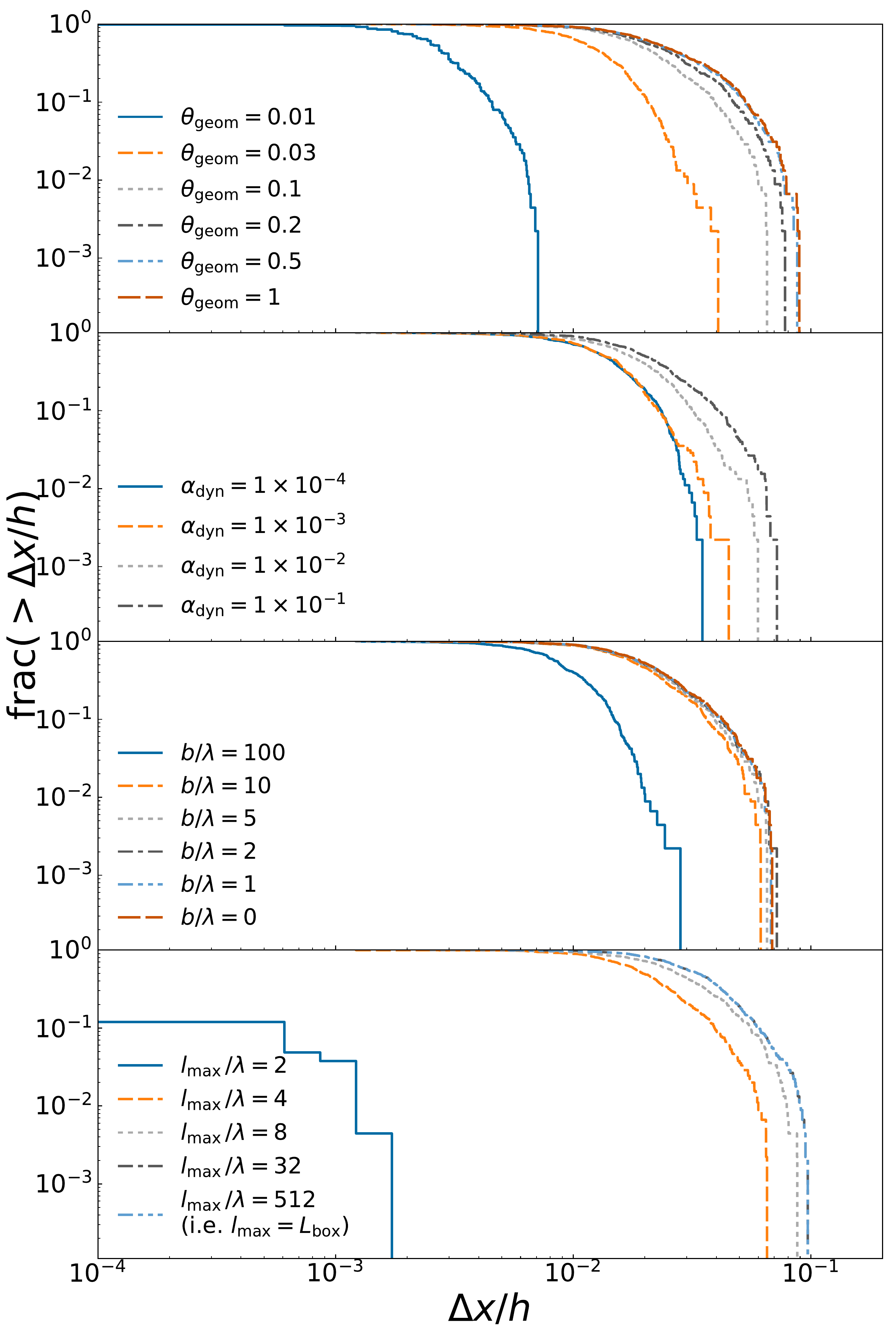}
    \caption{Similar to Fig.~\ref{fig:displacement_cumul}, but showing only the distribution at $z=0.03$. Each panel refers to a simulation set where only one parameter has been changed, namely from top to bottom: the opening angle $\tgeom$ used in the geometrical merging criterion, the parameter $\adyn$ used in the dynamical de-refinement condition, the size of the buffer zone $b$ (in units of the mean interparticle distance $\lambda$), and the maximum length of de-refined tree nodes $\lmax$ (effectively enforcing a minimum background resolution). Each line corresponds to a different value of the parameter investigated, with the most conservative choice on top and the least one on the bottom of the legend. 
    }
    \label{fig:displacement_cumul_allparams}
\end{figure}

As done in Sec.~\ref{sec:accuracy}, we match the particles in twin simulations, enabling a \textit{particle by particle} direct comparison of their positions. The results of varying, separately, each one of the four parameters listed above are shown in Fig.~\ref{fig:displacement_cumul_allparams}. In each panel, all parameters but one are kept fixed to the values used in the \textsc{large} runs (\ie $\tgeom = 0.1$, $b=5$, $\lmax = 128 \, \hMpc$). The values investigated are reported in the legend accompanying each panel, ordered with the most conservative value at the top and the most aggressive one at the bottom). 
The data presented in the Figure are complemented by Table~\ref{table:particle_gain_all} where we list for each simulation the ratio between the initial and final particle number $N_\mathrm{p}^\mathrm{fin} / N_\mathrm{p}^\mathrm{init}$, as well as the reduction in computational time $T_\mathrm{wc}^\mathrm{DZS} / T_\mathrm{wc}^\mathrm{STD}$. Note that, as discussed in Sec.~\ref{sec:performance}, the setup of these simulations is such that DZS is somewhat restrained. Hence, the performance boost should be interpreted as a lower limit for what can be achieved in more realistic settings (\ie employing higher resolution or a more complete set of physical processes simulated). We will address this issue in a future work, where we will apply our DZS algorithm in a realistic simulation setting. 

It should be noted that the effects of the DZS parameters investigated here are intertwined with the ones due to the tree-opening criterion used in the force computation. In particular, whenever the latter introduces errors larger than the one added by DZS, the differences in the integration accuracy are suppressed. Hence, the results we present in the following should be considered indicative of the trends but not necessarily independent of other simulation parameters. From our tests, some features appear clear:
\begin{itemize}
    \item The accuracy is only weakly dependent on the geometrical opening angle (top panel of Fig.~\ref{fig:displacement_cumul_allparams}), with the exception of the lowest (i.e. most conservative) value $\tgeom = 0.01$. This is the case because for such a low value most of the simulation volume remains at the original resolution for most of the time, in contrast with the situation occurring for larger $\tgeom$. This can be seen also in Table~\ref{table:particle_gain_all}, where the run with $\tgeom = 0.01$ at $z=0$ still contains 32\% of the initial particles, in contrast with $\tgeom = 0.1$ (or even $\tgeom = 0.05$, not reported for simplicity) where only $\sim 1$\% of the original particles are left. In fact, an intermediate behavior is observed for $\tgeom = 0.03$. Finally, we note here that almost all the values of $\tgeom$ investigated are smaller than those typically employed in tree codes (where $\theta \gtrsim 0.7$), but this choice is dictated by the fact that DZS permanently changes the simulations and, for any given particle, such modifications are anisotropic. Both reasons call for a more conservative value of $\theta$ to be employed.
    
    \item When the dynamical de-refinement criterion is employed, the integration accuracy depends less, but more steadily, on the $\adyn$ parameter value.
    
    \item The buffer size has no effect on the accuracy, with the exception of $b=100$. As for $\tgeom = 0.01$, the reason is that such a large buffer size (corresponding to $1/5$ of the box size) prevents many particles outside of the lightcone from being de-refined ($7.3$\% of the initial particles are left at $z=0$, in contrast to $1.7$\% for $b=10$).
    
    \item The background resolution appears to be an important parameter in setting the integration accuracy. In fact, limiting the de-refined node size to be at most $\lmax / \lambda = 2$ (corresponding to an \textit{average} mass ratio of $8$ between low- and high-resolution particles) yields an excellent accuracy (with a maximum $\Delta x / h \approx 10^{-3}$ at $z=0$). Allowing for an extra `level' of de-refinement (\ie $\lmax / \lambda = 4$, corresponding to a mass ratio of $64$) deteriorates the accuracy by more than one order of magnitude. Further increasing $\lmax / \lambda$ only slightly worsens the accuracy (up to a factor of $2$ using $\lmax = \Lbox$, \ie $\lmax / \lambda = 512$).
\end{itemize}

\begin{table}
    \centering
        \begin{tabular}{lcc}
        \hline
        $\tgeom$ & $N_\mathrm{p}^\mathrm{fin} / N_\mathrm{p}^\mathrm{init}$ & $T_\mathrm{wc}^\mathrm{DZS} / T_\mathrm{wc}^\mathrm{STD}$ \\
        \hline

        $0.01$ & $0.3180$ & $0.9196$ \\
        $0.03$ & $0.0343$ & $0.7042$ \\
        $0.1^\dagger$  & $0.0165$ & $0.6372$ \\
        $0.2$  & $0.0158$ & $0.6250$ \\
        $0.5$  & $0.0157$ & $0.6278$ \\
        $1.0$  & $0.0157$ & $0.6043$ \\
        \hline
        
        &&\\
        &&\\
        
        \hline
        $\adyn$ & $N_\mathrm{p}^\mathrm{fin} / N_\mathrm{p}^\mathrm{init}$ & $T_\mathrm{wc}^\mathrm{DZS} / T_\mathrm{wc}^\mathrm{STD}$ \\
        \hline

        $1 \times 10^{-4}$ & $0.0214$ & $0.7404$ \\
        $1 \times 10^{-3}$ & $0.0169$ & $0.6859$ \\
        $1 \times 10^{-2}$ & $0.0160$ & $0.6631$ \\
        $1 \times 10^{-1}$ & $0.0157$ & $0.6295$ \\
        \hline

        &&\\
        &&\\
        
        \hline
        $b$ & $N_\mathrm{p}^\mathrm{fin} / N_\mathrm{p}^\mathrm{init}$ & $T_\mathrm{wc}^\mathrm{DZS} / T_\mathrm{wc}^\mathrm{STD}$ \\
        \hline

        $100$ & $0.0730$ & $0.7982$ \\
        $10$  & $0.0170$ & $0.6443$ \\
        $5^\dagger$   & $0.0165$ & $0.6148$ \\
        $2$   & $0.0163$ & $0.6850$ \\
        $1$   & $0.0162$ & $0.6130$ \\
        $0$   & $0.0162$ & $0.6202$ \\
        \hline
        
        &&\\
        &&\\
        
        \hline
        $\lmax / \lambda$ & $N_\mathrm{p}^\mathrm{fin} / N_\mathrm{p}^\mathrm{init}$ & $T_\mathrm{wc}^\mathrm{DZS} / T_\mathrm{wc}^\mathrm{STD}$ \\
        \hline

        $2$   & $0.7622$ & $0.9863$ \\
        $4^\dagger$   & $0.0165$ & $0.6291$ \\
        $8$   & $0.0031$ & $0.6026$ \\
        $42$  & $0.0016$ & $0.6267$ \\
        $512$ & $0.0016$ & $0.6040$ \\
        \hline
        \multicolumn{3}{l}{$^\dagger$ This is the standard value of the parameter in this Paper.}
        
        \end{tabular}
    \caption{Particle ($N_\mathrm{p}^\mathrm{fin} / N_\mathrm{p}^\mathrm{init}$) and time ($T_\mathrm{wc}^\mathrm{DZS} / T_\mathrm{wc}^\mathrm{STD}$) reduction for different parameters choices in the DZS algorithm. Each table refers to run of the \textsc{large} simulation where a single different parameter has been changed with respect to the fiducial configuration used throughout the paper (highlighted in the Table by $^\dagger$), namely from top to bottom: geometrical opening angle $\tgeom$, parameter $\adyn$ used in the dynamical opening criterion, buffer size $b$, and maximum length $\lmax$ allowed for a node to be de-refined (effectively enforcing a minimum background resolution). 
    Note that in our fiducial configuration, the maximum theoretical gain in the particle number is $N_\mathrm{p}^\mathrm{fin} / N_\mathrm{p}^\mathrm{init} = 0.0157$ because of the background resolution enforced.}
    \label{table:particle_gain_all}
\end{table}

\section{Performance}
\label{sec:performance}
After assessing the accuracy of the DZS algorithm, we now move to assess its performance. We do so first using our test suite, and then extrapolating its results to realistic next-generation simulations. Before delving into this topic, let us note that throughout the paper we employ a very simple numerical setup. Namely, we only simulate DM particles and ignore baryons. 
We also reduce to the bare minimum the operations performed on the simulated particles and the output processing (\eg we do not compute the particle potential, we do not run an on-the-fly halo finder, etc.). All these operations typically become more expensive toward low redshift (either because of structure collapse or because of the time-sampling required is normally finer), where DZS has the largest impact. Hence, all the efficiency boosts presented in the following should be considered as conservative estimates of what can be obtained for more realistic code configurations.

\subsection{Performance in our test suite}
\label{sec:performance_test}

\begin{figure}
    \centering
    \includegraphics[width=\columnwidth]{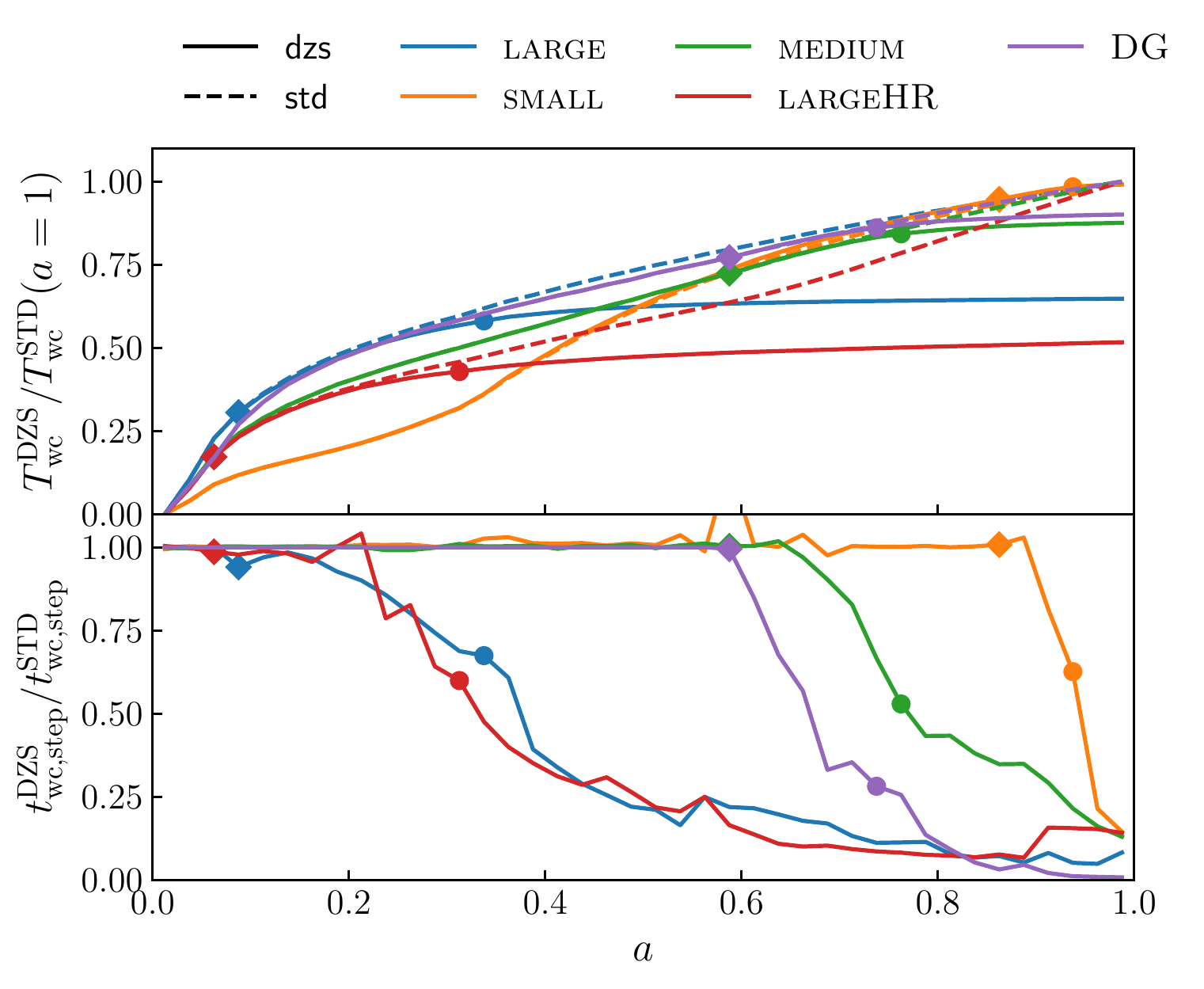}
    \caption{Top: Cumulative simulation wall-clock time --~normalised to its final value~-- as a function of expansion factor for the \dzs (solid line) and \std (dashed line) runs of the Llinares suite and the \textsc{DG} re-run. Bottom: Ratio of wall-clock time per timestep in each pair of twin runs. In both panels, diamond symbols mark the time when particles are de-refined for the first time. Similarly, circles indicate when the total number of particles in the \textit{dzs} simulation drops to half its initial value.}
    \label{fig:cpu_time_ratio}
\end{figure}

We start by investigating the performance of our implementation of DZS in the test suite summarised in Table~\ref{table:test_L17}. In the top panel of Fig.~\ref{fig:cpu_time_ratio} we show the wall-clock time $T_\mathrm{wc} (a)$ required by pairs of twin simulations as a function of the cosmological scale factor $a$, using solid lines for the \dzs runs and dashed ones for the \std runs. In order to compare runs with different characteristics (\eg number of particles, resolution, number of MPI tasks, etc.) we normalize each 
curve 
to the total time required by the \std run, which is always 
the largest among the two. 
For all simulations pairs, the time required by the two runs is exactly the same until the DZS algorithm starts merging together particles, reducing the workload. At first, this reduction is marginal, 
but it quickly becomes prominent as soon as
the number of particles is significantly reduced.

This is shown by the diamond symbols (circles) in Fig.~\ref{fig:cpu_time_ratio}, which mark the expansion factor when one (half) of the particles has been de-refined. The exact timing of these events strongly depends on the simulation box size and on the parameters used for DZS. 
Finally, note that in the case of the \textsc{DG} re-run we have started the simulation pair from the most advanced snapshot available \textit{before} the lightcone enters the simulation box, \ie $a \approx 0.6$, in order to save computational resources. Before that time, only a single run has been carried out, as DZS could not have any impact. 

The impact of DZS on the global simulation time also depends on the evolution of the computational cost. 
For instance, it is well-known that, at sufficiently-high resolution, matter clustering tends to increase the wall-clock time required per simulated time toward low redshift. However, its impact depends on many algorithmic and physical parameters. Therefore, in order to provide a more general measure of performance, we show in the bottom panel of Fig.~\ref{fig:cpu_time_ratio} the ratio between the time employed by the \dzs and \std runs to advance the simulation by a (fixed) interval in expansion factor\footnote{We choose this approach since DZS modifies the particle masses and distribution, and therefore the simulation timesteps differ, preventing a step-to-step comparison.} $\Delta a \approx 0.025$ (labeled as `step' in the Figure for the sake of brevity). As soon as the number of de-refined particles starts to be significant (of order 30\% of the total), the wall-clock time per step drops dramatically, quickly reaching $< 10$\% of its initial value.

We note here that when the simulation resolution increases, low-redshift timesteps typically become increasingly expensive. Hence, the late-time boost in computational efficiency provided by DZS is expected to become more and more effective when the resolution is increased (as we will detail more quantitatively in Fig.~\ref{fig:gain_map} below), rendering DZS a promising tool for the next generation of large cosmological simulations. A first indication of this behaviour can be seen by comparing, in the top panel of Fig.~\ref{fig:cpu_time_ratio}, the \textsc{large} and \textsc{largeHR} simulation pairs (which have the same box size, \ie the same onset of the DZS de-refinement procedure, and differ only for their resolution). The latter shows a significantly higher efficiency than the former, despite their per-timestep boost (bottom panel of the same Figure) being very similar. We will provide an estimate of the potential efficiency boost for such kind of simulations in Sec.~\ref{sec:estimation}. However, increased resolution entails an increased complexity in the numerical treatment (\eg increased number of computing tasks and communication) which can partially reduce the efficiency, especially when the particle distribution is greatly altered during the simulation, as it is the case for DZS. Hence, in the following we analyse the effect of DZS on the code workload and memory balance.

\subsubsection{Memory and workload balance}
\label{sec:imbalance}
\begin{figure}
    \centering
    \includegraphics[width=\columnwidth]{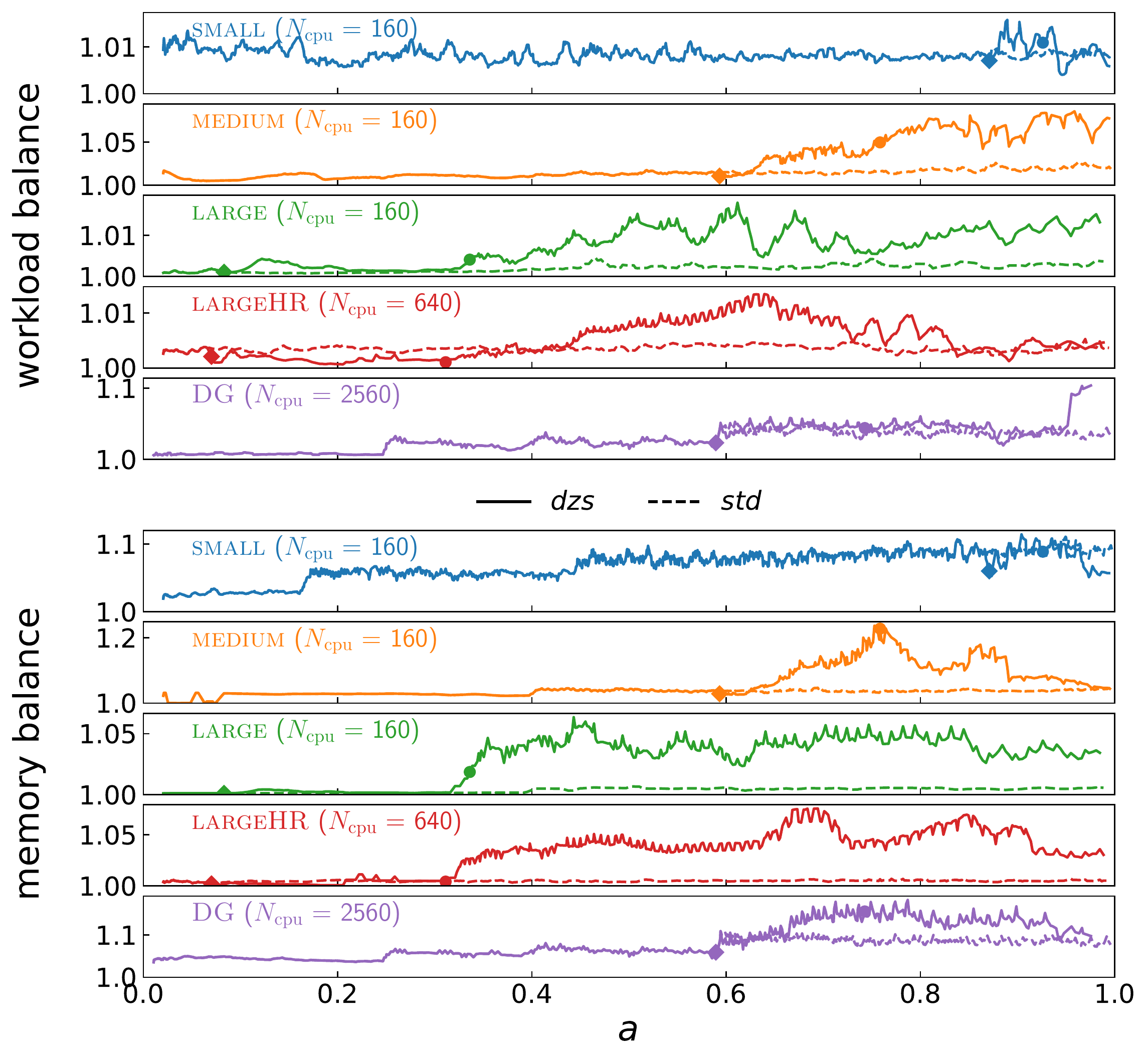}
    \caption{Workload balance (top panels) and memory balance (bottom panels) evolution with simulated expansion factor for all the runs listed in Table~\ref{table:test_L17} (with the exception of \textsc{tiny}). Solid curves refer to the \dzs run, while dashed lines to \std runs. As in Fig.~\ref{fig:showcase}, we mark with a diamond symbol the expansion factor when the first de-refinement takes place, and with a circle when the total particle number drops to half its initial value. The curves are smoothed using a running average with window size equal to $5$ timesteps.}
    \label{fig:imbalance}
\end{figure}

Fig.~\ref{fig:imbalance} shows the workload (top panels) and memory (bottom panels) balance for our twin simulations, defined as the maximum across all tasks of, respectively, the time spent on the tree walk and the memory allocated, divided by their average value. Hence a value of $1$ corresponds to an ideal balance in both cases. (Note that for visualization purposes, the panels cover slightly different ranges along the vertical axis). Each panel shows results concerning one pair of the twin simulations listed in Table~\ref{table:test_L17}, using a solid (dashed) line for the \dzs (\std) run. 
We also report in the top left corner of each panel the number of computing tasks $N_\mathrm{cpu}$ used for the twin runs, since it plays an important role in determining the imbalance.
As expected, as soon as DZS begins to systematically de-refine particles the two curves diverge and the \dzs run has an increased imbalance. This is an unavoidable effect since particles are de-refined based on their position, upon which depends also their location on computing tasks. However, the native domain decomposition algorithm of the \gadget code, based on a space-filling Peano-Hilbert curve, appears able to efficiently re-adjust the particle distribution and keep the imbalance under control. In fact, in all the \dzs runs the workload balance always remains below $1.1$, and in most cases below $1.02$. Similarly, the memory balance is always below $1.1$ in all but two cases. Additionally, the imbalance level reached entirely depends on the details of the algorithm employed to keep it under control, and therefore can largely change among different codes. Hence, the number reported here could be improved upon by different simulation codes.

It is interesting to note how in the \textsc{small} run, the effect of DZS on the workload and memory balance is negligible, as a consequence of the matter clustering (more evident in this simulation pair since it reaches the largest mass resolution) imposing a more substantial toll on the simulation balance than the de-refinement effect of DZS. Indeed, the memory balance starts very close to its ideal value at $a \sim 0$ but increases to its final value well before the DZS algorithm begins de-refining particles ($a \gtrsim 0.8$, marked with a diamond in all panels of Fig.~\ref{fig:imbalance}).

The simulation performing the worst in terms of both workload and memory balance is \textsc{medium}, which combines a relatively large box (\ie allowing DZS to be active for long time) and a decent mass resolution, enabling the formation of a sizeable number cosmic structures. Similarly, the \textsc{DG} re-run, which has a comparable size but better mass resolution, reaches an analogous level of memory imbalance, but its workload balance tracks the one of the \std run for most of the simulation time, surging only at the very end when few particles are left.

It should be noted that memory imbalance alone is somewhat less of an issue when DZS is employed. Typically, the main concern in having an imbalanced memory load is that it forces the usage of more computing tasks than needed. In fact, each one of them needs to have sufficient memory to store the maximum amount of information any task can be assigned, which is much larger than the average if the imbalance is high. Beyond the increased time spent by the code in communication, this is particularly problematic whenever the amount of computational resources is limited. However, when such imbalance is triggered by DZS, the amount of information stored in the simulation decreases at the same time. Hence, even the increased imbalance is always associated with a \textit{smaller} total memory requirement.

Although our tests show that \gadget is able to limit the imbalance to few per cent, a level not expected to affect the code performance, it is difficult to precisely predict the workload imbalance reached by large-box high-resolution simulation. In any case, even if the workload balance remained always under control during the whole numerical integration, the effect of the DZS technique necessarily implies that the computational requirements of a simulation will change during the run, as DZS progressively reduces the amount of information stored and, hence, diminishes the required total memory. Adjusting the computational setup to the evolving needs of a simulation can prevent resources from being wasted because of almost-empty computing tasks which spend a significant fraction of time communicating or waiting, a feature particularly useful when the amount of resources is limited.
For these reasons, we have included in our \gadget implementation of DZS a series of conditions that force the code to stop and produce an output whenever the imbalance is too large. This will allow users to modify the number of computing tasks the code is run on (and/or other numerical parameters) in order to dynamically optimize the allocation of resources required for continuing the simulation. In particular, we force the code to stop whenever one of the following conditions is met:
\begin{itemize}
    \item the workload balance exceeds a user-defined threshold;
    \item the ratio between the current and initial number of particles per task $\mathcal{R} (t) \equiv N_\mathrm{part} (t) / N_\mathrm{part}^\mathrm{init}$ is below a user-defined threshold;
    \item once the code has reached its run-time limit, and $\mathcal{R} (t)$ is predicted (using a linear extrapolation) to fall below the user-defined threshold within a user-defined time from the next simulation re-start.
\end{itemize}
The latter condition is devised to avoid a simulation from being forced to stop soon after it re-started. In this case, it is more efficient to slightly relax the halt criterion and enforce the special output earlier in order to allow the simulations parameters to be adjusted before requesting a new (and more appropriate) allocation of resources.

\subsection{Performance estimation for next-generation simulations}
\label{sec:estimation}
The simulations employed in this work were devised to allow the exploration of different parameter choices for the DZS algorithm, as well as different numerical setups. For this reason, a compromise had to be made between the mass resolution achieved and the size of the simulation. However, the expectation is that DZS will become increasingly more efficient with increasing simulation volume \textit{and} mass resolution. While we plan to carry out simulations in this regime in a future work, we want to estimate here the performance boost that can be expected from DZS for an arbitrary simulation setup.

In order to have realistic estimations, 
we need to take into account the effect of small-scale clustering of dark matter particles, which significantly increases the computation time and is particularly prominent at low redshift. We do so by running a simulation with small box but high particle resolution (compared to the other simulations employed in this work), namely $\Lbox = 128 \, \hMpc$ and $N_\mathrm{part} = 512^3$ (dubbed \textsc{tiny}), and using its normalized wall-clock time evolution $T_\mathrm{wc} (a) / T_\mathrm{wc} (1)$ as a template for a simulation with larger box size but equal mass resolution. Notice that the small box size implies that large-scale density fluctuations are missing, overall suppressing the densest regions. At the same time, the box is too small to include the most massive structures in the Universe. Both effects imply that such simulation underestimates the late-time matter clustering, and hence the performance boost estimated using it is likely smaller than what can be achieved in a realistic setup. 

It can be seen in Fig.~\ref{fig:cpu_time_ratio} that shortly after DZS has de-refined a significant number of particles, the computational cost of the simulation grows only very mildly, contributing only to a small part of the final wall-clock time. To first order, we can approximate this by assuming that a run with DZS enabled has the same cost as one without it until some critical expansion factor $a_\mathrm{DZS}$, and $0$ afterwards, \ie:
\begin{equation}
    \frac{T_\mathrm{wc}^\mathrm{STD}(a=1)}{T_\mathrm{wc}^\mathrm{DZS}(a=1)} = \frac{T_\mathrm{wc}^\mathrm{STD}(a=1)}{T_\mathrm{wc}^\mathrm{STD}(a_\mathrm{DZS})} ~.
\end{equation}
Note that this quantity is the reciprocal of the one plotted in the top panel of Fig.~\ref{fig:cpu_time_ratio}.
In order to be conservative in our estimate, we can use as critical expansion factor the one at which the lightcone volume is half of the simulation one (approximately corresponding to the circles in Fig.~\ref{fig:cpu_time_ratio}). Using \textsc{tiny} as a template for $T_\mathrm{wc}^\mathrm{STD} (a)$ 
we can estimate the performance boost for a simulation similar to the Euclid flagship one, \ie $\Lbox = 4000 \, \hMpc$ and $N_\mathrm{part} = 4 \times 10^{12}$ \citep{pkdgrav-euclid-flagship} to be \textit{at least}
\begin{equation}
\frac{T_\mathrm{wc}^\mathrm{STD}(a=1)}{T_\mathrm{wc}^\mathrm{DZS}(a=1)} =\frac{T_\mathrm{wc}^\mathrm{STD}(a=1)}{T_\mathrm{wc}^\mathrm{STD}(\approx 0.12)} \approx 3 ~.
\end{equation}
We stress again that this is likely to be a severe under-estimation of the actual achievable speedup.

In a similar way, we can estimate the performance boost for a simulation with the same characteristics as the DEUS `full Universe' run \citep{DEUS} to be $\gtrsim 10$. However, it must be noted that such simulation is very peculiar, as by construction it encloses the lightcone from the very beginning. This represents a configuration that we have not tested, so it may present a slightly different behaviour, hence altering the estimated performance boost.

\begin{figure}
    \centering
    \includegraphics[width=\columnwidth]{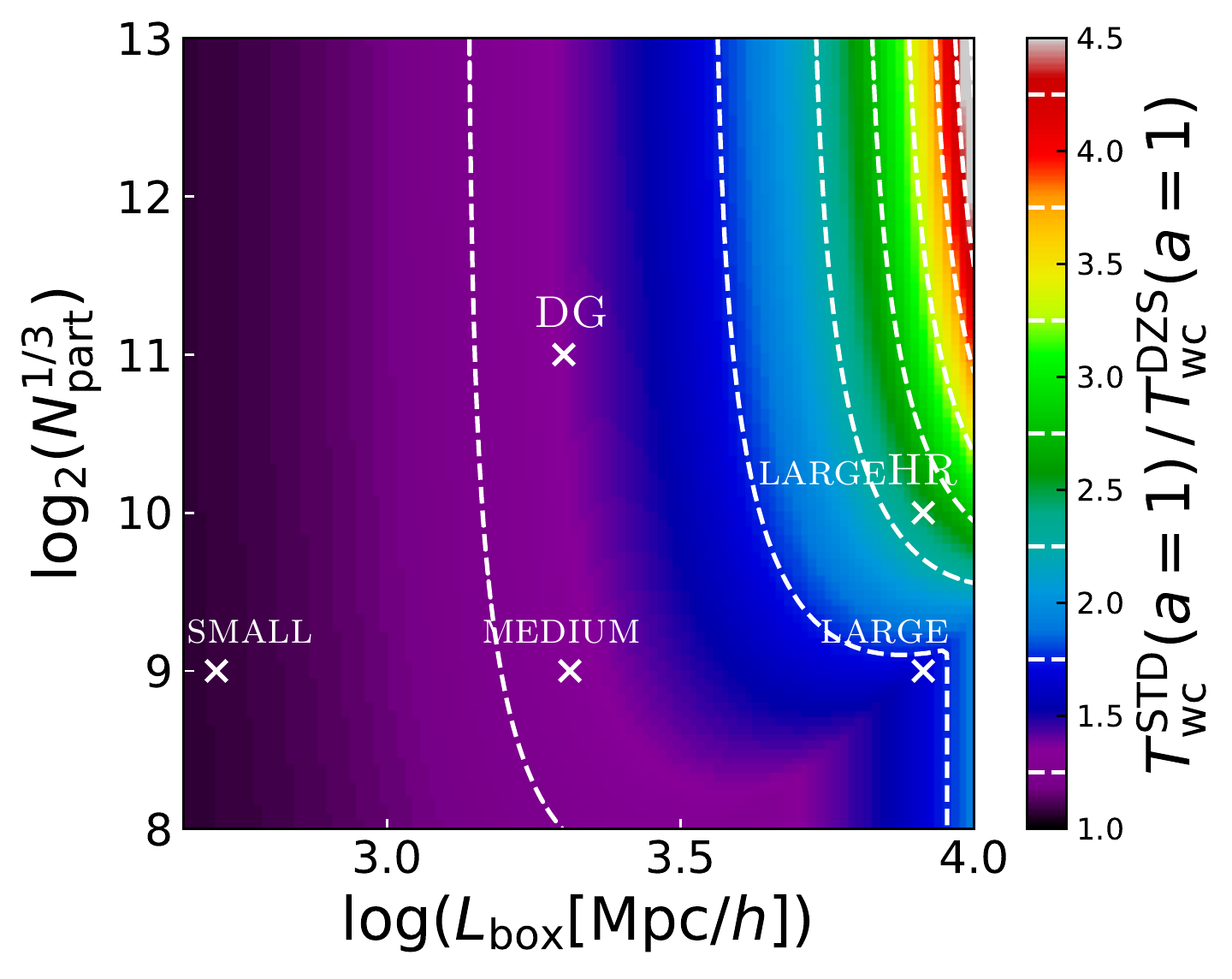}
    \caption{Predicted speedup enabled by DZS as a function of box size ($\Lbox$) and number of particles($N_\mathrm{part}$). Crosses show where our test simulations lie in this plane (note that \textsc{tiny} is outside of the plotted range). This is a qualitative prediction, see text for caveats. Dashed lines show contours of equal gain, as indicated in the colorbar.}
    \label{fig:gain_map}
\end{figure}

In order to give a qualitative overview of the speed-up enabled by DZS, we show in Fig.~\ref{fig:gain_map} the predicted ratio of wall-clock times for a simulation without and with DZS (the reciprocal of the quantity plotted on the vertical axis of the top panel of Fig.~\ref{fig:cpu_time_ratio}) as a function of box size ($\Lbox$) and number of particles($N_\mathrm{part}$). In order to compute this quantity, we employ the same approach used above for the Euclid flagship simulation. However, we need to account for the different shape of $T_\mathrm{wc}^\mathrm{STD} (a)$ 
as a function of mass resolution (due to matter clustering at low redshift having a larger impact at higher mass resolution). We do so in an empirical way, noticing that in our lowest-resolution simulation (\textsc{large}) such curve is well approximated by $a^{0.4}$, while in the highest-resolution one (\textsc{tiny}) it is approximately a straight line (\ie $\approx a$). Hence, we approximate the \textit{shape} of $T_\mathrm{wc}^\mathrm{STD} (a)$ 
with a curve $a^r$, \ie
$T_\mathrm{wc}^\mathrm{STD}(a)/T_\mathrm{wc}^\mathrm{STD}(a=1) \approx a^r$, where $r$ is linearly interpolated between the value $r=0.4$ at our coarsest mass resolution and $r=1$ at the highest one. For better mass resolutions, the value of $r$ is extrapolated, while for coarser resolutions it is kept fixed at $r=0.4$, since the clustering is expected to be negligible in such regime. 
Hence, we end up with the following estimate:
\begin{align}
    &\frac{T_\mathrm{wc}^\mathrm{STD}(a=1)}{T_\mathrm{wc}^\mathrm{DZS}(a=1)} \approx \frac{T_\mathrm{wc}^\mathrm{STD}(a=1)}{T_\mathrm{wc}^\mathrm{STD}(a_\mathrm{DZS})} \approx a_\mathrm{DZS}^{- \max \left( 0.4, 0.4 + \frac{1 - 0.4}{m_\textsc{small} - m_\textsc{large}} (m - m_\textsc{large}) \right)} \\
    &a_\mathrm{DZS} : \frac{4}{3} \pi \Rlc(a_\mathrm{DZS})^3 = \frac{1}{2} \Lbox^3 ~,
\end{align}
where $m$ in the first equation is the dark matter mass resolution of the simulation (computed from $\Lbox$ and $N_\mathrm{part}$ assuming our fiducial cosmology).
In Fig.~\ref{fig:gain_map} we mark with crosses the locations of our test suite runs in such plane. As predicted, DZS brings the largest benefit for large box size and number of particles. It should be noted that for $\log_2(N_\mathrm{part}^{1/3}) \gtrsim 10$ the estimated performance boost is approximately independent of $N_\mathrm{part}$. This may be partially due to our extrapolation method. 
Alternatively, it could simply indicate that the reduction in the number of particles (driven by $\Lbox$) is dominant over the effect of matter clustering due to increasing resolution.

\section{Conclusions}
\label{sec:conclusions}
In this paper we have presented a new method - named Dynamic Zoom Simulations, or DZS in short - devised to speed up cosmological simulations while preserving their accuracy in the production of lightcone outputs. Our approach combines seamlessly the relativistic notion of lightcone with the Newtonian nature of typical cosmological simulations in a way that is fully compatible with the latter, preserves the periodicity properties, and does not require any substantial change in the code structure. \\

The DZS algorithm de-refines particles outside of the observer's lightcone (see Fig.~\ref{fig:showcase}) in order to reduce the total computational time while preserving the large scale density field. While this can, in principle, be done in different ways, we employ a tree-based approach. This ensures scalability, allows a smooth transition between the high- and low-resolution regions, and integrates effortlessly with treePM codes. The latter is of particular importance since: (i) it ensures virtually no overhead in the computational time; (ii) it renders the implementation of DZS into existing treePM codes relatively simple (as the necessary infrastructure is already in place), and (iii) enables the exploitation of DZS together with the broad range of features that have been implemented in popular simulation codes over the years.

We have tested the accuracy and performance of this approach by implementing it in the popular \gadget code, and list in the following our main results:

\begin{itemize}
    \item The DZS algorithm is able to reduce the wall-clock time per time step by more than an order of magnitude once the lightcone radius is a fraction of the box size (Fig.~\ref{fig:cpu_time_ratio}). This translates into a reduction of the total computation time of order 50\% for the simplest configuration we have tested, and likely much more for higher-resolution or more complex (\ie involving more simulated physics and/or outputs) simulations.
    
    \item The de-refinement of particles frees up progressively more memory during the simulation. Hence, individual computing tasks progressively store less information, and spend larger fractions of time in communication. This situation is particularly problematic for large simulations and/or with constrained total resources available. For this reason, we have devised and implemented a mechanism that optimizes the halt and re-start of a simulation in order to fine tune the computational resources assigned to the run. Additionally, this could also help reducing the small imbalances in the memory and workload distribution across different tasks (Fig.~\ref{fig:imbalance}) that can arise as a consequence of the particle de-refinement.
    
    \item The comparison between simulations with and without DZS shows that the lightcone halo mass function (Fig.~\ref{fig:hmf}), the sky-projected matter lightcone and its angular power spectrum (Fig.~\ref{fig:healpix_diff}), and the 3D matter lightcone (Fig.~\ref{fig:3d_lightcone_skymap_combined}) are unchanged to the $\sim 0.1$\% level.
    
    \item The integration accuracy is mostly unchanged by the DZS algorithm, while the simulation time is significantly reduced. In our largest-box highest-resolution test (\ie where the effect of DZS is maximal) the particle positions (across the entire simulated time) in twin runs with and without DZS differ by at most few times the mean inter-particle distance $h$. More than $99$\% of the particles are displaced less than $0.2 \, h$ (Fig.~\ref{fig:displacement_cumul}), while the total runtime is halved.
    
    \item The integration accuracy can be tweaked thanks to a number of parameters specific to DZS, with a consequent change in the wall-clock time. By investigating a broad region of the parameter space (Fig.~\ref{fig:displacement_cumul_allparams} and Table~\ref{table:particle_gain_all}) we provide some guidance on how to optimize the DZS parameters in order to achieve the optimal trade-off between accuracy and performance boost for the application of choice.
    
    \item Combining a high-resolution small-box simulation with coarse-resolution large-box ones, we provide an estimation of the efficiency boost that can be expected when employing DZS for any combination of box size and particle number (\ie mass resolution, Fig.~\ref{fig:gain_map}). In the region of the $\Lbox$-$N_\mathrm{part}$ space that new-generation simulations are starting to explore, we forcast DZS to reduce the total simulation \textit{at least} by  a factor of a few in the case of DM-only simulation, and likely much more so for runs including baryons (although DZS is currently implemented only for DM-only simulations).
\end{itemize}

To summarise, we have presented a novel approach to the production of lightcones in cosmological simulations. Our Dynamic Zoom Simulation --~or DZS~-- approach de-refines particles once they are not needed anymore for the lightcone production,  in order to reduce the workload of the simulation while simultaneously maintaining its output accuracy and the large-scale gravitational field. This allows to 
significantly improve the performance with respect to equal simulations carried out with the traditional approach. A number of parameters control the behaviour of DZS, enabling a customization of the accuracy and speed-up reached through its use. In DZS, traditional time-slice output have reduced resolution outside of the lightcone (\ie in the entire box at $z=0$). Our implementation allows the user to enforce a minimum volume to be kept at maximum resolution until the end of the simulation, in order to have a high-resolution time-slice of a portion of the simulation box even at $z=0$.

Overall, our results show that DZS is a promising approach for the development of large high-resolution simulations, as the ones required to model and interpret the wealth of high-quality data that upcoming surveys will provide.

\section*{Acknowledgements}
The main idea behind this work came up during a coffee break discussion at a \textit{Euclid} Cosmological Simulations Working Group meeting that took place in 2017 in Barcelona. The authors are therefore grateful to all the people that took part in that stimulating (yet very informal) discussion: Pablo Fosalba, Romain Teyssier, Joachim Stadel, Claudio Llinares, Joop Schaye, Aurel Schneider, Carlo Giocoli, Linda Blot.
We thank Volker Springel and Claudio Llinares for useful comments on the manuscript. We additionally thank Volker Springel for making \gadget available to us. We are thankful to the community developing and maintaining software packages extensively used in our work, namely: matplotlib \citep{matplotlib}, numpy \citep{numpy}, scipy \citep{scipy}, cmasher \citep{cmasher}, healpy \citep{healpy}.
 
\section*{Data availability}
The data underlying this article will be shared on reasonable request to the corresponding author.

\bibliographystyle{mnras}
\bibliography{biblio}{}

\begin{thebibliography}{}
\makeatletter
\relax
\def\mn@urlcharsother{\let\do\@makeother \do\$\do\&\do\#\do\^\do\_\do\%\do\~}
\def\mn@doi{\begingroup\mn@urlcharsother \@ifnextchar [ {\mn@doi@}
  {\mn@doi@[]}}
\def\mn@doi@[#1]#2{\def\@tempa{#1}\ifx\@tempa\@empty \href
  {http://dx.doi.org/#2} {doi:#2}\else \href {http://dx.doi.org/#2} {#1}\fi
  \endgroup}
\def\mn@eprint#1#2{\mn@eprint@#1:#2::\@nil}
\def\mn@eprint@arXiv#1{\href {http://arxiv.org/abs/#1} {{\tt arXiv:#1}}}
\def\mn@eprint@dblp#1{\href {http://dblp.uni-trier.de/rec/bibtex/#1.xml}
  {dblp:#1}}
\def\mn@eprint@#1:#2:#3:#4\@nil{\def\@tempa {#1}\def\@tempb {#2}\def\@tempc
  {#3}\ifx \@tempc \@empty \let \@tempc \@tempb \let \@tempb \@tempa \fi \ifx
  \@tempb \@empty \def\@tempb {arXiv}\fi \@ifundefined
  {mn@eprint@\@tempb}{\@tempb:\@tempc}{\expandafter \expandafter \csname
  mn@eprint@\@tempb\endcsname \expandafter{\@tempc}}}

\bibitem[\protect\citeauthoryear{{Alimi} et~al.,}{{Alimi} et~al.}{2012}]{DEUS}
{Alimi} J.-M.,  et~al., 2012, arXiv e-prints, \href
  {https://ui.adsabs.harvard.edu/abs/2012arXiv1206.2838A} {p. arXiv:1206.2838}

\bibitem[\protect\citeauthoryear{{Arnold}, {Fosalba}, {Springel}, {Puchwein}
  \& {Blot}}{{Arnold} et~al.}{2019}]{MICE-lc}
{Arnold} C.,  {Fosalba} P.,  {Springel} V.,  {Puchwein} E.,   {Blot} L.,  2019,
  \mn@doi [\mnras] {10.1093/mnras/sty3044}, \href
  {https://ui.adsabs.harvard.edu/abs/2019MNRAS.483..790A} {483, 790}

\bibitem[\protect\citeauthoryear{{Bagla}}{{Bagla}}{2002}]{Bagla2002}
{Bagla} J.~S.,  2002, \mn@doi [Journal of Astrophysics and Astronomy]
  {10.1007/BF02702282}, \href
  {https://ui.adsabs.harvard.edu/abs/2002JApA...23..185B} {23, 185}

\bibitem[\protect\citeauthoryear{{Bagla} \& {Ray}}{{Bagla} \&
  {Ray}}{2003}]{Bagla+2003}
{Bagla} J.~S.,  {Ray} S.,  2003, \mn@doi [\na] {10.1016/S1384-1076(03)00056-3},
  \href {https://ui.adsabs.harvard.edu/abs/2003NewA....8..665B} {8, 665}

\bibitem[\protect\citeauthoryear{{Baldi}}{{Baldi}}{2012}]{Codecs}
{Baldi} M.,  2012, \mn@doi [\mnras] {10.1111/j.1365-2966.2012.20675.x}, \href
  {https://ui.adsabs.harvard.edu/abs/2012MNRAS.422.1028B} {422, 1028}

\bibitem[\protect\citeauthoryear{{Baldi}, {Pettorino}, {Robbers}  \&
  {Springel}}{{Baldi} et~al.}{2010}]{C-gadget}
{Baldi} M.,  {Pettorino} V.,  {Robbers} G.,   {Springel} V.,  2010, \mn@doi
  [\mnras] {10.1111/j.1365-2966.2009.15987.x}, \href
  {https://ui.adsabs.harvard.edu/abs/2010MNRAS.403.1684B} {403, 1684}

\bibitem[\protect\citeauthoryear{{Bode}, {Ostriker}  \& {Xu}}{{Bode}
  et~al.}{2000}]{Bode+2000}
{Bode} P.,  {Ostriker} J.~P.,   {Xu} G.,  2000, \mn@doi [\apjs]
  {10.1086/313398}, \href
  {https://ui.adsabs.harvard.edu/abs/2000ApJS..128..561B} {128, 561}

\bibitem[\protect\citeauthoryear{{Bonafede}, {Dolag}, {Stasyszyn}, {Murante}
  \& {Borgani}}{{Bonafede} et~al.}{2011}]{Bonafede+2011}
{Bonafede} A.,  {Dolag} K.,  {Stasyszyn} F.,  {Murante} G.,   {Borgani} S.,
  2011, \mn@doi [\mnras] {10.1111/j.1365-2966.2011.19523.x}, \href
  {https://ui.adsabs.harvard.edu/abs/2011MNRAS.418.2234B} {418, 2234}

\bibitem[\protect\citeauthoryear{Clarke, Glendinning  \& Hempel}{Clarke
  et~al.}{1994}]{MPI}
Clarke L.,  Glendinning I.,   Hempel R.,  1994, in Decker K.~M.,  Rehmann
  R.~M.,  eds, Programming Environments for Massively Parallel Distributed
  Systems. Birkh{\"a}user Basel, Basel, pp 213--218

\bibitem[\protect\citeauthoryear{{DESI Collaboration} et~al.,}{{DESI
  Collaboration} et~al.}{2016a}]{DESI1}
{DESI Collaboration} et~al., 2016a, arXiv e-prints, \href
  {https://ui.adsabs.harvard.edu/abs/2016arXiv161100036D} {p. arXiv:1611.00036}

\bibitem[\protect\citeauthoryear{{DESI Collaboration} et~al.,}{{DESI
  Collaboration} et~al.}{2016b}]{DESI2}
{DESI Collaboration} et~al., 2016b, arXiv e-prints, \href
  {https://ui.adsabs.harvard.edu/abs/2016arXiv161100037D} {p. arXiv:1611.00037}

\bibitem[\protect\citeauthoryear{{Dai}, {Feng}  \& {Seljak}}{{Dai}
  et~al.}{2018}]{COLAbaryons}
{Dai} B.,  {Feng} Y.,   {Seljak} U.,  2018, \mn@doi [\jcap]
  {10.1088/1475-7516/2018/11/009}, \href
  {https://ui.adsabs.harvard.edu/abs/2018JCAP...11..009D} {2018, 009}

\bibitem[\protect\citeauthoryear{{Dolag} \& {Stasyszyn}}{{Dolag} \&
  {Stasyszyn}}{2009}]{Dolag+Stasyszyn2009}
{Dolag} K.,  {Stasyszyn} F.,  2009, \mn@doi [\mnras]
  {10.1111/j.1365-2966.2009.15181.x}, \href
  {https://ui.adsabs.harvard.edu/abs/2009MNRAS.398.1678D} {398, 1678}

\bibitem[\protect\citeauthoryear{{Evrard} et~al.,}{{Evrard}
  et~al.}{2002}]{Evrard+2002}
{Evrard} A.~E.,  et~al., 2002, \mn@doi [\apj] {10.1086/340551}, \href
  {https://ui.adsabs.harvard.edu/abs/2002ApJ...573....7E} {573, 7}

\bibitem[\protect\citeauthoryear{{Fosalba}, {Gazta{\~n}aga}, {Castand er}  \&
  {Manera}}{{Fosalba} et~al.}{2008}]{OnionUniverse}
{Fosalba} P.,  {Gazta{\~n}aga} E.,  {Castand er} F.~J.,   {Manera} M.,  2008,
  \mn@doi [\mnras] {10.1111/j.1365-2966.2008.13910.x}, \href
  {https://ui.adsabs.harvard.edu/abs/2008MNRAS.391..435F} {391, 435}

\bibitem[\protect\citeauthoryear{{G{\'o}rski}, {Hivon}, {Banday}, {Wandelt},
  {Hansen}, {Reinecke}  \& {Bartelmann}}{{G{\'o}rski} et~al.}{2005}]{healpix}
{G{\'o}rski} K.~M.,  {Hivon} E.,  {Banday} A.~J.,  {Wandelt} B.~D.,  {Hansen}
  F.~K.,  {Reinecke} M.,   {Bartelmann} M.,  2005, \mn@doi [\apj]
  {10.1086/427976}, \href {http://adsabs.harvard.edu/abs/2005ApJ...622..759G}
  {622, 759}

\bibitem[\protect\citeauthoryear{{Hollowed}}{{Hollowed}}{2019}]{Hollowed2019}
{Hollowed} J.,  2019, arXiv e-prints, \href
  {https://ui.adsabs.harvard.edu/abs/2019arXiv190608355H} {p. arXiv:1906.08355}

\bibitem[\protect\citeauthoryear{Hunter}{Hunter}{2007}]{matplotlib}
Hunter J.~D.,  2007, Computing In Science \& Engineering, 9, 90

\bibitem[\protect\citeauthoryear{{Ivezi{\'c}} et~al.,}{{Ivezi{\'c}}
  et~al.}{2019}]{LSST}
{Ivezi{\'c}} {\v Z}.,  et~al., 2019, \mn@doi [\apj] {10.3847/1538-4357/ab042c},
  \href {http://adsabs.harvard.edu/abs/2019ApJ...873..111I} {873, 111}

\bibitem[\protect\citeauthoryear{Jones, Oliphant, Peterson  et~al.}{Jones
  et~al.}{2001}]{scipy}
Jones E.,  Oliphant T.,  Peterson P.,   et~al., 2001, {SciPy}: Open source
  scientific tools for {Python}, \url {http://www.scipy.org/}

\bibitem[\protect\citeauthoryear{{Klypin} \& {Shandarin}}{{Klypin} \&
  {Shandarin}}{1983}]{Klypin+1983}
{Klypin} A.~A.,  {Shandarin} S.~F.,  1983, \mn@doi [\mnras]
  {10.1093/mnras/204.3.891}, \href
  {https://ui.adsabs.harvard.edu/abs/1983MNRAS.204..891K} {204, 891}

\bibitem[\protect\citeauthoryear{{Laureijs} et~al.,}{{Laureijs}
  et~al.}{2011}]{EUCLID}
{Laureijs} R.,  et~al., 2011, arXiv e-prints, \href
  {https://ui.adsabs.harvard.edu/abs/2011arXiv1110.3193L} {p. arXiv:1110.3193}

\bibitem[\protect\citeauthoryear{{Leclercq}, {Faure}, {Lavaux}, {Wand elt},
  {Jaffe}, {Heavens}, {Percival}  \& {No{\^u}s}}{{Leclercq}
  et~al.}{2020}]{Leclercq+2020}
{Leclercq} F.,  {Faure} B.,  {Lavaux} G.,  {Wand elt} B.~D.,  {Jaffe} A.~H.,
  {Heavens} A.~F.,  {Percival} W.~J.,   {No{\^u}s} C.,  2020, arXiv e-prints,
  \href {https://ui.adsabs.harvard.edu/abs/2020arXiv200304925L} {p.
  arXiv:2003.04925}

\bibitem[\protect\citeauthoryear{{Llinares}}{{Llinares}}{2017}]{Llinares17}
{Llinares} C.,  2017, arXiv e-prints, \href
  {https://ui.adsabs.harvard.edu/abs/2017arXiv170904703L} {p. arXiv:1709.04703}

\bibitem[\protect\citeauthoryear{{Nelson} et~al.,}{{Nelson}
  et~al.}{2019}]{TNG50-Dylan}
{Nelson} D.,  et~al., 2019, \mn@doi [\mnras] {10.1093/mnras/stz2306}, \href
  {https://ui.adsabs.harvard.edu/abs/2019MNRAS.490.3234N} {490, 3234}

\bibitem[\protect\citeauthoryear{{Nori} \& {Baldi}}{{Nori} \&
  {Baldi}}{2018}]{Ax-gadget}
{Nori} M.,  {Baldi} M.,  2018, \mn@doi [\mnras] {10.1093/mnras/sty1224}, \href
  {https://ui.adsabs.harvard.edu/abs/2018MNRAS.478.3935N} {478, 3935}

\bibitem[\protect\citeauthoryear{{Pawlik} \& {Schaye}}{{Pawlik} \&
  {Schaye}}{2008}]{TRAPHIC}
{Pawlik} A.~H.,  {Schaye} J.,  2008, \mn@doi [\mnras]
  {10.1111/j.1365-2966.2008.13601.x}, \href
  {https://ui.adsabs.harvard.edu/abs/2008MNRAS.389..651P} {389, 651}

\bibitem[\protect\citeauthoryear{{Petkova} \& {Springel}}{{Petkova} \&
  {Springel}}{2009}]{Petkova&Springel2009}
{Petkova} M.,  {Springel} V.,  2009, \mn@doi [\mnras]
  {10.1111/j.1365-2966.2009.14843.x}, \href
  {https://ui.adsabs.harvard.edu/abs/2009MNRAS.396.1383P} {396, 1383}

\bibitem[\protect\citeauthoryear{{Pillepich} et~al.,}{{Pillepich}
  et~al.}{2019}]{TNG50-Annalisa}
{Pillepich} A.,  et~al., 2019, \mn@doi [\mnras] {10.1093/mnras/stz2338}, \href
  {https://ui.adsabs.harvard.edu/abs/2019MNRAS.490.3196P} {490, 3196}

\bibitem[\protect\citeauthoryear{{Planck Collaboration} et~al.,}{{Planck
  Collaboration} et~al.}{2016}]{Planck2015_cosmology}
{Planck Collaboration} et~al., 2016, \mn@doi [\aap]
  {10.1051/0004-6361/201525830}, \href
  {https://ui.adsabs.harvard.edu/abs/2016A&A...594A..13P} {594, A13}

\bibitem[\protect\citeauthoryear{{Potter}, {Stadel}  \& {Teyssier}}{{Potter}
  et~al.}{2017}]{pkdgrav-euclid-flagship}
{Potter} D.,  {Stadel} J.,   {Teyssier} R.,  2017, \mn@doi [Computational
  Astrophysics and Cosmology] {10.1186/s40668-017-0021-1}, \href
  {https://ui.adsabs.harvard.edu/abs/2017ComAC...4....2P} {4, 2}

\bibitem[\protect\citeauthoryear{{Puchwein}, {Baldi}  \& {Springel}}{{Puchwein}
  et~al.}{2013}]{MG-gadget}
{Puchwein} E.,  {Baldi} M.,   {Springel} V.,  2013, \mn@doi [\mnras]
  {10.1093/mnras/stt1575}, \href
  {https://ui.adsabs.harvard.edu/abs/2013MNRAS.436..348P} {436, 348}

\bibitem[\protect\citeauthoryear{Spergel et~al.,}{Spergel
  et~al.}{2015}]{WFIRST}
Spergel D.,  et~al., 2015, Wide-Field InfrarRed Survey Telescope-Astrophysics
  Focused Telescope Assets WFIRST-AFTA 2015 Report (\mn@eprint {arXiv}
  {1503.03757})

\bibitem[\protect\citeauthoryear{{Springel}}{{Springel}}{2005}]{gadget2}
{Springel} V.,  2005, \mn@doi [\mnras] {10.1111/j.1365-2966.2005.09655.x},
  \href {http://adsabs.harvard.edu/abs/2005MNRAS.364.1105S} {364, 1105}

\bibitem[\protect\citeauthoryear{Springel, Yoshida  \& White}{Springel
  et~al.}{2001}]{gadget1}
Springel V.,  Yoshida N.,   White S.~D.,  2001, \mn@doi [New Astron.]
  {10.1016/S1384-1076(01)00042-2}, 6, 79

\bibitem[\protect\citeauthoryear{{Tassev}, {Zaldarriaga}  \&
  {Eisenstein}}{{Tassev} et~al.}{2013}]{COLA}
{Tassev} S.,  {Zaldarriaga} M.,   {Eisenstein} D.~J.,  2013, \mn@doi [\jcap]
  {10.1088/1475-7516/2013/06/036}, \href
  {https://ui.adsabs.harvard.edu/abs/2013JCAP...06..036T} {2013, 036}

\bibitem[\protect\citeauthoryear{{Tassev}, {Eisenstein}, {Wand elt}  \&
  {Zaldarriaga}}{{Tassev} et~al.}{2015}]{sCOLA}
{Tassev} S.,  {Eisenstein} D.~J.,  {Wand elt} B.~D.,   {Zaldarriaga} M.,  2015,
  arXiv e-prints, \href {https://ui.adsabs.harvard.edu/abs/2015arXiv150207751T}
  {p. arXiv:1502.07751}

\bibitem[\protect\citeauthoryear{{Viel}, {Haehnelt}  \& {Springel}}{{Viel}
  et~al.}{2010}]{Viel+2010}
{Viel} M.,  {Haehnelt} M.~G.,   {Springel} V.,  2010, \mn@doi [\jcap]
  {10.1088/1475-7516/2010/06/015}, \href
  {https://ui.adsabs.harvard.edu/abs/2010JCAP...06..015V} {2010, 015}

\bibitem[\protect\citeauthoryear{Walt, Colbert  \& Varoquaux}{Walt
  et~al.}{2011}]{numpy}
Walt S. v.~d.,  Colbert S.~C.,   Varoquaux G.,  2011, Computing in Science \&
  Engineering, 13, 22

\bibitem[\protect\citeauthoryear{{White}, {Frenk}  \& {Davis}}{{White}
  et~al.}{1983}]{White+1983}
{White} S.~D.~M.,  {Frenk} C.~S.,   {Davis} M.,  1983, \mn@doi [\apjl]
  {10.1086/184139}, \href
  {https://ui.adsabs.harvard.edu/abs/1983ApJ...274L...1W} {274, L1}

\bibitem[\protect\citeauthoryear{{Xu}}{{Xu}}{1995}]{Xu1995}
{Xu} G.,  1995, \mn@doi [\apjs] {10.1086/192166}, \href
  {https://ui.adsabs.harvard.edu/abs/1995ApJS...98..355X} {98, 355}

\bibitem[\protect\citeauthoryear{Zonca, Singer, Lenz, Reinecke, Rosset, Hivon
  \& Gorski}{Zonca et~al.}{2019}]{healpy}
Zonca A.,  Singer L.,  Lenz D.,  Reinecke M.,  Rosset C.,  Hivon E.,   Gorski
  K.,  2019, \mn@doi [Journal of Open Source Software] {10.21105/joss.01298},
  4, 1298

\bibitem[\protect\citeauthoryear{{van der Velden}}{{van der
  Velden}}{2020}]{cmasher}
{van der Velden} E.,  2020, \mn@doi [The Journal of Open Source Software]
  {10.21105/joss.02004}, \href
  {https://ui.adsabs.harvard.edu/abs/2020JOSS....5.2004V} {5, 2004}

\makeatother
\end{thebibliography}

\bsp	
\label{lastpage}
\end{document}